\documentclass{aa}  
\usepackage{graphicx}
\usepackage{txfonts}
\usepackage{aalongtable,lscape}
\newcommand{\kms}{\ensuremath{\mathrm{km\,s}^{-1}}}
\newcommand{\vsini}{\ensuremath{v_\mathrm{e} \sin i}}

\newcommand{\te}{\ensuremath{T_{\mathrm{eff}}}}
\newcommand{\llo}{\ensuremath{L/L_\odot}}
\newcommand{\logl}{\ensuremath{\log(L/L_\odot)}}
\newcommand{\mmo}{\ensuremath{M/M_\odot}}

\newcommand{\bz}{\ensuremath{\langle B_z \rangle}}
\newcommand{\mo}{\ensuremath{M_\odot}}
\newcommand{\brms}{\ensuremath{B_\mathrm{rms}}}

\begin{document}
   \title{Searching for links between magnetic fields and stellar evolution}
   \subtitle{II. The evolution of magnetic fields as revealed by 
                 observations of Ap stars in open clusters and associations
                 }
   \author{J.~D.~Landstreet \inst{1}
       \and 
        S.~Bagnulo          \inst{2, 3}
       \and
        V.~Andretta         \inst{4}
       \and 
        L. Fossati          \inst{5}
       \and
        E.~Mason            \inst{2}
       \and
        J.~Silaj            \inst{1}
       \and
        G.~A.~Wade          \inst{6}
        }
\offprints{J.~Landstreet}
\institute{Physics \& Astronomy Department,
           The University of Western Ontario,
           London, Ontario, Canada N6A 3K7. \\
           \email{jlandstr@astro.uwo.ca, jsilaj@astro.utoronto.ca}
           \and
           European Southern Observatory,
           Alonso de Cordova 3107, Vitacura,
           Santiago, Chile \\
           \email{emason@eso.org}
           \and
           Armagh Observatory, College Hill,
           Armagh, BT61 9DG, Northern Ireland \\
           \email{sba@arm.ac.uk}
           \and
           INAF - Osservatorio Astronomico di Capodimonte, 
           salita Moiariello 16, 80131 Napoli, Italy.
           \email{andretta@na.astro.it}
           \and
           Institut f\"{u}r Astronomie, Wien Universitaet,
           Tuerkenschanzstr.\ 17, A-1180 Wien, Austria. 
           \email{fossati@astro.univie.ac.at} 
           \and
           Department of Physics,
           Royal Military College of Canada,
           P.O. Box 17000, Station `Forces'
           Kingston, Ontario, Canada K7K 7B4. \\
           \email{Gregg.Wade@rmc.ca}
           }

   \date{Received: 1 Jan 2009 ; accepted: ever hopeful }

 
   \abstract
{The evolution of magnetic fields in Ap stars during the main
sequence phase is presently mostly unconstrained by observation
because of the difficulty of assigning accurate ages to known field Ap
stars. }
{We are carrying out a large survey of magnetic fields in cluster Ap
  stars with the goal of obtaining a sample of these stars with
  well-determined ages. In this paper we analyse the information
  available from the survey as it currently stands.}
{We select from the available observational sample the stars that are probably
  (1) cluster or association members and (2) magnetic Ap stars. For
  the stars in this subsample we determine the fundamental parameters
  \te, \llo, and \mmo. With these data and the cluster ages we
  assign both absolute age and fractional age (the
  fraction of the main sequence lifetime completed). For this purpose we 
  have derived new bolometric corrections for Ap stars.}
{Magnetic fields are present at the surfaces of Ap stars from the ZAMS
  to the TAMS. Statistically for the stars with $M > 3 \mo$ the fields
  decline with advancing age approximately as expected from flux
  conservation together with increased stellar radius, or perhaps even
  faster than this rate, on a time scale of about $3\,10^7$ yr. In
  contrast, lower mass stars show no compelling evidence for field
  decrease even on a timescale of several times $10^8$ yr.}
{Study of magnetic cluster stars is now a powerful tool for obtaining
  constraints on evolution of Ap stars through the main
  sequence. Enlarging the sample of known cluster magnetic stars, and
  obtaining more precise RMS fields, will help to clarify the results
  obtained so far. Further field observations are in progress. }

\titlerunning{The evolution of magnetic fields in Ap stars}
\authorrunning{J.~D.~Landstreet, S.~Bagnulo, V.~Andretta, et al.}
   \maketitle
%

\section{Introduction}

A small fraction, of the order of 10\,\%, of the main sequence A and B
stars show obvious signs of anomalous atmospheric chemical abundances
in their spectra. The distinctive surface abundances of these stars
have never found a satisfactory explanation as the result of formation
in a chemically anomalous region of the galaxy, or as the product of
internal nuclear evolution or of surface nuclear reactions. Instead,
it is now generally thought that these stars have undergone chemical
fractionation as a result of competition between gravitationally
driven diffusion of trace elements, radiative levitation of some of
these elements, and interaction of the separation process with mass
loss and with mixing processes such as convection and turbulence
(Michaud \cite{Michaud70}; Babel \cite{Babel92}; Richer et al.
\cite{Richeretal00}).

One of the major groups of these main sequence stars with anomalous
atmospheric abundances is the ``magnetic peculiar A'' stars, which are
commonly referred to as Ap stars. These stars all have large-scale (global)
magnetic fields with strengths in the range of a few hundred to about
30\,000 G. Their distinctive chemical anomalies appear to vary fairly
systematically with \te; the cooler ones (Ap SrCrEu stars with $7000
\leq \te \leq 10000$ K) show strong overabundances of Sr, Cr, and rare
earths, while hotter ones (Ap Si and Ap He-wk stars, with $10000 \leq
\te \leq 15000$) typically show strong Si overabundance and He
underabundance. The hottest Ap stars (the He-str stars with $\te
\sim 20\,000$ K) have overabundant He. The broad features of which
elements are expected to be over or underabundant as a function of
\te\ are understood -- to some extent -- as a result of the competing
processes mentioned above.

The magnetic Ap stars are also usually periodically variable, in light,
spectrum, and magnetic field. The periods range from about half a day
to decades without any obvious difference in the phenomena of
variation, and the periods observed are inversely correlated with
\vsini. It is clear from these facts that the period of variation must
be the rotation period. This variability implies that the
field and the chemical abundances are non-uniformly distributed over
the surface of an Ap star. The rotation axis is not an axis of symmetry, 
so that the observer sees a configuration that changes as the star rotates.

Much is now understood about these magnetic Ap stars. Simple models of
field structure indicate that almost all magnetic Ap stars have
roughly dipolar field structures. The axis of (approximate) symmetry
of the field is usually oblique to the rotation axis, so that the
line-of-sight component of the field varies roughly sinusoidally with
time (e.g. Mestel \& Landstreet \cite{MestelLandstreet05}).  Often
several chemical elements are distributed quite non-uniformly over the
surface, and low-resolution models (e.g. Strasser, Landstreet \&
Mathys \cite{Strasseretal01}) of the distributions are available for a
number of Ap stars. It is now becoming possible to deduce quite
detailed maps of both field structure and element distribution for a
few stars for which the variations have been observed
spectroscopically in all four Stokes parameters (Kochukhov, Bagnulo,
Wade et al.  \cite{kea04}; Kochukhov \cite{Koc04}).

The lack of any energetically significant convection zone near the
surface of all but the coolest Ap stars, and the absence of a direct
correlation between rotation rate and field strength (as is found in
solar-type stars, see Mestel \& Landstreet \cite{MestelLandstreet05})
has led to general agreement that the magnetic fields observed are
probably fossil fields retained from an earlier evolution stage, a
situation which is possible because of the extremely long decay times.

Numerical calculations have suggested how a fossil magnetic field
might evolve with time during the long main sequence phase as a
consequence of ohmic decay, changes in stellar structure, and internal
circulation currents (Moss \cite{Moss01}), but the relationship of
such computations to observed stars is still very unclear.
Braithwaite \& Spruit (\cite{BraithwaiteSpruit04}) and Braithwaite \&
Nordlund (\cite{BraithwaiteNordlund06}) have shown that the
combination of the Cowling instability with rotation and meridional
circulation may lead to major reorganisation of the internal flux.
Similarly, theoretical computations suggest how atmospheric chemical
peculiarities might develop and evolve with time through the main
sequence phase, but so far these computations have not succeeded in
reproducing observed abundance patterns in detail (Babel \& Michaud
\cite{BabelMichaud91}). Observational data on fields and abundances in
a sample of Ap stars for which the absolute age and fraction of the
main sequence evolution completed (we will call this the ``fractional
age'') is known would be extremely valuable to test theoretical ideas
and provide clues to important processes.

At present, we have few constraints from observations about the
evolution of either fields or atmospheric chemistry with time before
or through the main sequence phase.  Ap stars have somewhat anomalous,
and variable, energy distributions, which make it difficult to place
these stars in the HR diagram precisely enough to obtain accurate ages
for individual stars from comparison with theoretical evolution
tracks. The problem is compounded by the fact that the atmospheric
chemistry does not readily reveal the bulk stellar abundances, and so
we do not even know how to choose the metallicity used in computing
the evolution tracks which are to be compared with observations. As a
result, until recently very few Ap star ages could be securely
determined. Age determinations of field stars are barely accurate
enough to determine whether a given star is in the first half or
second half of its main sequence life. (This problem is discussed in
more detail by Bagnulo et al. \cite{Bagea06}, hereafter Paper I).

The problem of obtaining an accurate age for an Ap star is considerably
simpler if the star is a member of a cluster. Since the age of a
cluster can be determined typically to somewhat better than $\pm 0.2$ dex
(by isochrone fitting to the main sequence turnoff or to
low-mass stars evolving to the zero-age main sequence, or by studies
of Li depletion in low-mass stars), we can use the cluster age to
determine the fraction of the main sequence lifetime which has
elapsed.  For a star in a cluster whose age is small compared to
the star's main sequence lifetime, this evolutionary age is relatively
precise. On the other hand, age resolution is not very good for the
second half of a star's main sequence lifetime (see Paper I, and Sect.~5.1).

It has recently become practical to obtain magnetic measurements of a
statistically interesting number of stars in clusters.  In Paper I we
have reported the first large scale observational survey of magnetic
fields in peculiar (and normal) A and B stars in open clusters and
associations. This survey has been motivated by our desire to obtain
empirical information about the evolution of magnetic fields and
atmospheric chemistry in middle main sequence stars as these stars
evolve from the zero-age main sequence (ZAMS) to the terminal-age main
sequence (TAMS).

In the present paper, we draw some first conclusions from this
(ongoing) survey about the evolution of magnetic fields from ZAMS to
TAMS. In Sect.~\ref{Sect_Star_Selection} we will examine and select
all the stars that are probable cluster or association members and
probable Ap stars, and for which magnetic measurements are available
(either from our own survey or previous works). In
Sect.~\ref{Sect_Physical_Properties} we determine the fundamental
parameters \te\ and \llo\ for these stars.  Combining this information
with suitable evolution tracks, we derive stellar masses and
fractional ages for the stars of the sample in
Sect.~\ref{Sect_Age_Mass}. In Sect.~\ref{Sect_Character} we comment
on the characteristics of our sample (distributions of age, mass, and
magnetic field). Finally, in Sect.~\ref{Sect_Field_Evol}, we
consider what these data reveal about the evolution of surface fields
from the ZAMS to the TAMS. Sect~7 summarises our conclusions.

\section{Cluster membership of stars with magnetic data}
\label{Sect_Star_Selection}

In Paper I we presented new magnetic measurements of many candidate or
confirmed Ap stars that may be members of clusters or associations (as
well as a large number of measurements of probable cluster members
that are not Ap stars). From these data and from field measurements in
the literature, we have assembled a database of all probable Ap
cluster/association members for which usefully precise magnetic field
measurements are available. This database is presented in
Table~\ref{cl_memb_pec_test.tab}; the content will be explained below.
In this section we analyze more critically the evidence for cluster
membership, and for magnetic Ap nature, for the stars in
Table~\ref{cl_memb_pec_test.tab}. This will enable us to remove any
stars for which either membership or Ap nature is doubtful from the
stellar sample we finally analyse.

\subsection{Assessing cluster membership}

Cluster (or association) membership may be tested in several ways.
Traditionally, membership is tested by plotting a colour-magnitude
diagram for the cluster. Stars which lie too far above or below the
cluster main sequence are rejected as members. Previous studies (e.g.
North \cite{North93}; Maitzen, Schneider \& Weiss
\cite{Maitzenetal88}) have shown that Ap stars that are probable
cluster members lie reasonably close (in observational HR diagrams,
for example Johnson $V$ magnitude as a function of colour $B-V$) to
the main sequence of normal stars, although Ap stars sometimes deviate
from the normal main sequence by somewhat more than normal cluster
members.  In general, the stars in Table~\ref{cl_memb_pec_test.tab}
are sufficiently close to the main sequences of the clusters to which
they may belong that this criterion does not immediately eliminate any
of the stars in the table from membership. The HR diagram test of
membership will be applied again later in this paper when the
provisionally accepted cluster members are placed in the theoretical
(\llo\ {\it vs} \te) HR diagrams of their specific clusters.

Comparisons of measured parallaxes and proper motions of candidate
stars with the average values for the cluster provide powerful tests
of membership. The situation for data on both these properties has
improved dramatically as a result of the Hipparcos mission (ESA
\cite{esa97}). This mission led to measurements of parallaxes and
proper motions of about 120\,000 stars, with typical errors of the
order of 1 mas and 1 mas yr$^{-1}$, so that accurate ($\pm 25$\,\%)
parallaxes are available for tens of thousands of stars out to a
distance of about 300 pc.  Observations of a much larger number of
stars than that included in the Hipparcos parallax programme, using
the star mapper instrument on the satellite, led to the Tycho (or
Tycho-1) Catalogue (H\o g et al. \cite{hogea97}) of proper motions of
about 1 million stars, with accuracies of the order of 10 mas
yr$^{-1}$ (at $V = 9$) in both coordinates. The Tycho data were
combined with positions from the Astrographic Catalogue to produce the
Tycho Reference Catalogue (H\o g et al. \cite{hogea98}), with proper
motions of 990\,000 stars accurate to about 2.5 mas yr$^{-1}$, and an
extension of this project led to the Tycho-2 catalogue (H\o g et al.
\cite{hogea00a}, \cite{hogea00b}), which combines data from the
Hipparcos mission with those from many ground-based position
catalogues to provide proper motions for about 2.5 million stars with
typical proper motion standard errors of 2.5 mas yr$^{-1}$.

These data have been used by several groups for important studies of
membership in clusters and OB associations. Robichon et al.
(\cite{robichon99}) determined mean distances, mean proper motions,
and membership of Hipparcos parallax stars located near the centres of
18 clusters less than 500 pc away. Membership was derived from
assessment of position, parallax, proper motion, and photometry. The
work led to list of accepted members of each cluster; the absence of a
Hipparcos parallax star from one of these lists presumably means that
it was not accepted by Robichon et al. (\cite{robichon99}) as a member
of the corresponding cluster.

Baumgardt et al. (\cite{baumgardt00}) derived mean distances and
motions for 205 open clusters using Hipparcos parallax stars found in
these clusters. Membership of Hipparcos parallax stars in these
clusters was assessed using photometry, radial velocity, position, and
proper motion data from the ground and from Hipparcos, together with
some additional proper motions from the Tycho Reference
Catalogue. Baumgardt et al. report both an
overall membership assessment (member, possible member, non-member)
and a proper motion membership probability.

The Hipparcos data were used by de Zeeuw et al.
(\cite{dezeeuw99}) to assess membership of Hipparcos
parallax stars in nearby OB associations. Several of the associations
included in their study (Sco OB2, $\alpha$ Per, Collinder 121,
Trumpler 10, etc) are represented in our survey. 

The release of the Tycho-2 proper motion catalogue 
made possible further large-scale studies of
membership in open clusters. Dias, L\'epine and Alessi (\cite{dias01},
\cite{dias02}) have determined mean proper motions of more than 200
clusters, many of which were not included in the studies discussed
above, using in most cases several tens of stars per cluster. They have
derived membership probabilities based only on the proper motions,
which are tightly clustered for an open cluster but (usually) have a
considerably larger dispersion for field stars.

Kharchenko et al. (\cite{kharchenko05})
created a database of possible stellar members of some 520 clusters.
They use a combination of Tycho-2, Hipparcos, and ground-based data to
determine average cluster parameters (distance, reddening, proper
motions, etc).  Memberships of individual stars in their database are
assessed using all available information (position, photometry, proper motions,
etc), and then cluster ages are derived using the positions of the
brightest stars relative to theoretical isochrones. 

We have used the studies discussed above together with the actual
astrometric data to assess the probability that the stars of
Table~\ref{cl_memb_pec_test.tab} are actually cluster members.
Wherever possible we have adopted the consensus of the previous
membership studies. However, the individual determinations of
membership probability differ somewhat from one study to another, both
because a study may be based either on Hipparcos or Tycho-2 proper
motions, and because the stars selected as definite cluster members,
from which the cluster mean motions are deduced, varies from one study
to another.  Note in particular that Kharchenko et al.
(\cite{kharchenko05}) often assign membership probabilities based on
proper motions that are significantly smaller than those found in any
of the other studies (e.g. for \object{HD 21699} in the $\alpha$ Per cluster,
de Zeeuw et al. \cite{dezeeuw99} give $P = 0.84$ and Robichon et al.
\cite{robichon99} consider the star to be a definite member, but
Kharchenko et al. (\cite{kharchenko05}) give $P_\mu = 0.25$ as the
proper motion membership probability). We
generally consider that the Kharchenko et al. data support membership
if their proper motion membership probability is larger than about
0.15 (their 2--$\sigma$ limit).

In a few cases the literature conclusions on membership are
discordant, or no membership probability based on recent astrometry is
available. In these cases we have compared proper motions and
parallaxes, when available, to the cluster means, and assessed
membership ourselves. We consider a star to be a probable member if
both proper motions (and parallax if available) are less than $2.5
\sigma$ from the mean cluster values, and a definite member if all are
within $1 \sigma$ of the cluster means.  If no accurate astrometry is
available, but the star is within the bounds of the cluster and has
colour and magnitude consistent with membership, we have generally
assumed that the star is a probable member.

In the particular case of stars in Ori OB1a, OB1b, and OB1c, the
members of the association are not well separated in proper motion
from the field. We have adopted the criterion of Brown et al.
(\cite{brown99}) for proper motion membership, and have tested apparent
brightness and parallax for consistency with membership as far as
possible. However, membership of stars in these regions of Ori OB1 is 
uncertain (see discussion in de Zeeuw et al.
\cite{dezeeuw99}). In Ori OB1d, where background obscuration virtually
eliminates background stars, we can be more confident of membership.

The results of this exercise are reported in
Table~\ref{cl_memb_pec_test.tab} in the four columns under the general
heading ``cluster membership''. In these columns we report our
conclusions as to membership probability considering only parallax
(``$\pi$''), proper motions (``$\mu$''), location in the HR diagram
relative to the cluster isochrone (``phot'', see below), and overall
(``member?''), using a four-level scale of ``y'' = member, ``p'' =
probable member, ``?'' = probably non-member, and ``n'' = non-member.
The ``member?'' column also contains references to earlier membership
studies.  Notes at the bottom of Table~\ref{cl_memb_pec_test.tab},
signaled by ``*'' in the ``member?'' column, discuss particular
problems with deciding on cluster or association membership.

\subsection{Assessing spectral type and chemical peculiarity}

We next turn to methods to determine whether a particular star is an
Ap star of the type in which a magnetic field is almost always
detected when sufficiently precise field measurements are obtained
(Auri\`ere et al. \cite{Auretal04}). The reason that
this is important is that the field measurement errors achieved in
general in Paper I are only small enough to
detect fields in about half of the magnetic Ap stars we have observed,
and since we will want to examine the distribution of field strengths
in magnetic Ap stars as a function of age and mass, it is important
to consider both the Ap stars in which we detect a field and those in
which no field has yet been found.

Several criteria are available that can indicate Ap type spectra.
Spectral classification can provide an important test of Ap
characteristics. For many of the stars in our sample, one or more
classification spectra are available. Classification surveys of a
number of clusters with the goal of identifying Ap stars have been
reported by Hartoog (\cite{hartoog76}, \cite{hartoog77}) and Abt
(\cite{abt79}) and references therein. Spectroscopic studies of single
clusters have also yielded many spectroscopic classifications (these
may be found in the WEBDA database of open cluster data at
http://www.univie.ac.at/webda/, described by Mermilliod \& Paunzen
\cite{MerPau03}; on the SIMBAD database of stellar data at
http://cdsweb.u-strasbg.fr/; and in the compilation of data on Ap
stars in clusters by Renson \cite{renson92}). Unfortunately, since
the classification criteria for Ap spectra are not always unambiguous,
and may not be clearly visible in low dispersion, low signal-to-noise
classifications spectra, the available classifications of a star are
frequently contradictory as to whether the star is actually an Ap.

Another valuable criterion for Ap type is furnished by the $\Delta a$
photometry system developed by Maitzen and collaborators
(e.g., Maitzen \cite{Mai93}), and by the $Z$ index which
can be computed for stars for which Geneva photometry is available
(e. g. Cramer \& Maeder \cite{CraMae79}, \cite{CraMae80}; Hauck \& North
\cite{HauNor82}). Both of these photometric indices are sensitive to
the broad energy depression at about 5200~\AA\ which has been found to
be a robust indicator of Ap spectral type in roughly the temperature
range 8\,000 -- 14\,000 K (e.g., Kupka, Paunzen, \& Maitzen
\cite{Kupetal03}). Both these photometry systems provide a large
amount of information about cluster Ap stars; Maitzen's group
has surveyed a large number of clusters for Ap
stars, and Geneva data exist for many cluster stars
as well (see http://obswww.unige.ch/gcpd/). 

A third useful criterion of Ap type is the detection of periodic
photometric variability of the type frequently found in Ap stars. The
catalogues of Renson \& Catalano (\cite{RensonCatalano01}) provide
very useful summaries of the many particular studies found in the
literature. We have searched the Hipparcos catalogue (ESA
\cite{esa97}) for new photometric variables, although this provided
only a very small number of variables that had not already been
identified by ground-based photometry.

Finally, the definite detection of a magnetic field is a clear
indication of Ap classification. 

We have summarized our view as to whether the star is an Ap star in
the four columns under the general heading ``Ap star'' of
Table~\ref{cl_memb_pec_test.tab}. We have used the same scale adopted to
evaluate the membership (``y'', ``p'', ``?'', and ``n''; a blank space
denotes absence of information).  In the column labelled ``Sp'' we
report the conclusions obtained from the available spectral
classifications.  We have adopted the criterion that $\Delta a$
photometry definitely supports Ap classification if $\Delta a \geq
0.014$, and that Geneva photometry supports Ap classification if $Z
\leq -0.016$ (within the temperature range where photometric detection
of peculiarity is valid). The conclusions are given in column $\Delta
a\, ,\, Z$.  The degree of support for Ap classification from
photometric variability is summarized in the column labelled ``var''.
The column ``mag fld'' reports our assessment as to whether the star
has a detected field, and also provides references to field
measurements. From this column it will be clear that a large fraction
of all field measurements of open cluster stars have been carried out
in the course of our survey (Paper~I; measurements denoted by ``F'' 
in the ``mag fld'' column); the only substantial
previous surveys have been restricted to the Sco OB2 and Ori OB1
associations.  Finally, an overall assessment as to whether the
balance of evidence supports an Ap classification is provided in the
last column ``Ap?''  of Table~\ref{cl_memb_pec_test.tab}. Note that a
significant number of stars are given a ``y'' or ``p'' even though no
magnetic field has yet been detected in them.

\section{Determining the physical properties of cluster Ap stars}
\label{Sect_Physical_Properties}

With the analysis described in the previous Sect.\ we have identified
a subsample of about 90 stars that are definite or probable (``y'' or
``p'') cluster/association members, definite or probable Ap stars,
and for which at least one magnetic field measurement is available
(regardless of whether a field has been detected or not). 

For each star of this subsample we need to determine age and mass.
These will be calculated in Sect.~\ref{Sect_Age_Mass}, using suitable
evolutionary models, assuming that we know the cluster age, the star's
temperature and the star's luminosity. In this Sect.\ we describe how
we determine these physical properties. Furthermore, to each star we
associate a value representative of the star's field strength.

\subsection{Cluster ages}

Because our essential goal is to provide a sample of magnetic
measurements of Ap stars of reasonably well known ages, the
ages of the clusters to which the stars in our sample belong must be
critically discussed. Cluster ages for a large sample of clusters,
including essentially all those considered here, are available from
the WEBDA database (\cite{MerPau03}). Ages in this
database are derived from a survey of literature values carried out by
Loktin et al. (\cite{loktin01}). These ages are in
most cases based on isochrone fits to dereddened cluster main
sequences.

Another extensive study of ages has been carried out by Kharchenko et
al (\cite{kharchenko05}) based on databases of potential cluster
members generated from large observational resources such as the
Hipparcos and Tycho databases. The ages of Kharchenko et al
are derived by fitting isochrones individually to the most massive
main sequence stars found among their own selection of cluster
members, rather than by isochrone fitting to the upper end of the
cluster main sequence as a whole. The ages of Kharchenko et al. are
based in many cases on only one star per cluster. Nevertheless, these
ages are in good overall agreement with those of Loktin et al.
(\cite{loktin01}): Kharchenko et al. have shown that (excluding a
small number of clusters where significant differences in choices of
cluster members lead to substantially different ages) the ages from
these two sources exhibit an RMS difference in cluster age ($\log t$)
of about $\sigma = 0.20$ dex. Note that neither WEBDA nor Kharchenko
et al. provide uncertainties for the assigned cluster ages; one of our
main tasks is to estimate these uncertainties.

We have included these ages in our own database of possible cluster Ap
stars, together with many recent determinations from the literature.
After examining the available recent age determinations, we have
selected what seems to us to be an appropriate age for each cluster,
and assigned an uncertainty to this value. In assigning ages, we have
considered both accuracy claimed by individual age studies and the
overall agreement of various studies. For a number of the clusters of
interest, several age determinations are available, and in many
cases these are reasonably concordant. For such clusters we have
assigned an uncertainty which in the best cases is taken to be $\pm
0.1$\,dex. For clusters with fewer age determinations, or more
discordant ones, we have generally assigned an age and an uncertainty
which approximately cover the range of the best age determinations
within the range 0.1 -- 0.2 dex. For clusters for which no ages
seem to be available, we have mostly used the ages from WEBDA (those
of Loktin et al.  \cite{loktin01}), and assumed an uncertainty of $\pm
0.2$\,dex, in agreement with the overall concordance between their
results and those of Kharchenko et al. (\cite{kharchenko05}).

In addition, several of the probable Ap stars we have observed turn
out to be near the TAMS in the HR diagram. By requiring the cluster
isochrone to intersect the error oval of such a relatively evolved
star, in some cases we are able to constrain the cluster age more
precisely than by using ages obtained from the literature; we obtain
uncertainties that may be as small as 0.05 dex. In these cases we have
adopted the cluster ages and uncertainties from our isochrone fits.
The procedure adopted is discussed further in Sect.  4.

\subsection{Stellar effective temperatures}

\subsubsection{Photometry}

For stars for which Geneva photometry is available, we have used the
FORTRAN code described by K\"{u}nzli, North, Kurucz \& Nicolet
(\cite{kunzli97}), which is available from CDS, to determine a first
effective temperature based on the calibration for normal (metallicity
0) stars. This procedure requires supplying a colour excess
$E(B_2 - V_1)$. We have obtained this quantity from the Johnson colour
excess $E(B - V)$ (either for the cluster as a whole, as reported by
WEBDA or a particular study, or for the star individually after
de-reddening in the Johnson $U-B, B-V$ colour-colour diagram), using
the relationship $E(B_2 - V_1) \approx 0.88 E(B - V)$ (Hauck \& North
\cite{HauNor93}). When the colours of a star are near the boundary
between different methods of temperature determination (e.g. $XY$ or
$pTpG$) we have computed the temperature using both methods. The
effective temperature determined using normal star calibrations is
then corrected to the Ap stars temperature scale as described by Hauck
\& K\"{u}nzli (\cite{hauck96}). For stars classified as He-weak or
He-strong, we have followed the suggestion of Hauck \& North to adopt
the temperature produced by the K\"{u}nzli et al FORTRAN programme
without correction.

For stars for which Str\"{o}mgren $uvby\beta$ photometry is available,
we have computed an effective temperature using the FORTRAN code
``UVBYBETANEW'' of Napiwotzki, Sch\"{o}nberner \& Wenske
(\cite{napiwotzki93}), which is in turn based mainly on the
calibrations of Moon \& Dworetsky (\cite{moon85}). When the star has
colours near the boundary of different calibrations (for example near
A0), we have computed the temperature using both calibrations. These
raw temperatures have been corrected for Ap stars as described by
St\c{e}pie\'n \& Dominiczak (\cite{stepien89}). Again, we have not
corrected temperatures of He-weak or He-strong stars from the values
produced by the programme.

If only Johnson $UBV$ photometry is available, we have de-reddened the
star to the main sequence line in the $U-B, B-V$ colour-colour
diagram, and then corrected the value following the same prescription
St\c{e}pie\'n \& Dominiczak (\cite{stepien89}) use for Str\"{o}mgren
photometry.

\subsubsection{Adopted temperatures and uncertainties}

For most of the stars in our sample, both $uvby\beta$ and Geneva 
photometry are
available, and we have adopted a suitable average of the computed
temperatures. We find that temperature determination using the two
different kinds of photometry, and sometimes more than one method of
temperature determination, have a scatter of the order of 2--300
K. Since both methods have been calibrated using mostly the same
``fundamental'' stars, this is certainly a lower bound to the actual
uncertainty of the results. 

The global temperature calibration for Ap stars is still not
very satisfactory. {\em No} Ap stars have a fundamental calibration
(based on integrated flux and angular diameter measurements); all
calibrations are based on corrections to the normal star calibrations
using a small sample of stars for which effective temperatures have
been determined using the infrared flux method (e.g. Lanz \cite{Lan85}; 
Megessier \cite{Meg88}; Monier \cite{monier92}), fitting
energy distributions (e.g. Adelman et al. \cite{Adeetal95}), or 
the reasonable but uncertain correction
method used by St\c{e}pie\'n \& Dominiczak (\cite{stepien89}), who
estimate the uncertainty of their individual values of \te\ to be
6--700 K. Furthermore, the accuracy of these corrections has recently
been called into question by the results of Khan \& Shulyak
(\cite{khan06}), who have calculated model atmospheres with enhanced
metal content (so far generally scaled solar abundances) {\em and} the
effects of a magnetic field. They find that, even with a rather large
field and 10x solar abundances, the Paschen continuum slope is very
similar to that of a solar composition, non-magnetic star, and that
the calibrations for {\em normal} stars recover approximately the
assumed effective temperature.

Finally, there are certainly substantial star-to-star variations in
atmospheric composition, which make any calibration suitable only on
average, and not exact for any particular star. This is clearly seen
in the scatter shown in Figs. 9 and 10 of Napiwotzki et al.
(\cite{napiwotzki93}), where individual calibration points are shown
along with the mean calibration curves adopted.

Considering the internal uncertainties in determining the effective
temperature of a magnetic Ap star, and the further global
uncertainties, we adopt a uniform uncertainty of $\pm 0.02$\,dex
(roughly $\pm 500$ K) for our temperatures, and even this value may be
somewhat optimistic.

\subsection{Stellar luminosities}

\subsubsection{Cluster distances}

In the case of the nearest clusters (out
to about 300 pc) accurate distances are best derived using 
Hipparcos parallaxes averaged over known
members. Distances obtained in this way generally have uncertainties
of about 0.2 mag or less in both absolute and apparent distance modulus
(cf Robichon et al. \cite{robichon99}). 

For more distant clusters, the most accurate distances are obtained by
isochrone fitting to dereddened cluster main sequences. For a number of 
nearby clusters, it is possible to compare distances obtained from 
isochrone fitting with those obtained from parallaxes (Robichon et 
al \cite{robichon99}). This comparison, and 
comparison of different determinations using isochrone fitting, allow
us to estimate that a typical uncertainty for the more distant clusters
is roughly 0.2 mag in distance modulus, similar to the value for 
the nearer clusters. In almost all cases the uncertainty
in the distance modulus is dominated by the uncertainty in the 
distance rather than that in the reddening.

We have
adopted the cluster or association distances for all of the stars in
our sample, rather than individual distances even when these are
available from parallax observations; on average the mean cluster
distances from parallaxes have standard errors that are smaller by a
factor of two or more than those of individual cluster stars. Of
course, for stars in more distant clusters, the best available distance
is the photometric distance modulus of the cluster as a whole.

\subsubsection{Bolometric corrections}

The results of Lanz (\cite{lanz84}) strongly suggest that the
bolometric corrections for magnetic Ap stars are different from those
of normal stars.  Since we have effective temperatures obtained from
several photometric systems, we need bolometric corrections suitable
for Ap stars as a function of effective temperature, rather than as a
function of a specific photometric index (as usually provided).
We have re-examined the available data to obtain accurate bolometric
corrections in the form we need.

For this purpose, we have collected all the magnetic Ap stars for
which integrated fluxes are available.  Bolometric corrections for
magnetic Ap stars have been reported by Lanz (\cite{lanz84}), while
St\c{e}pie\'n \& Dominiczak (\cite{stepien89}), and Monier
(\cite{monier92}) have provided further integrated fluxes. Almost all
the stars discussed by Lanz are rather hot (Si and He-pec stars), and
most have distances between 100 and 250 pc, so the observed fluxes are
significantly affected by interstellar absorption. Except for three
cool Ap stars in his sample, the fluxes reported by Lanz have been
corrected approximately for extinction.  The integrated fluxes
reported by St\c{e}pie\'n \& Dominiczak do not seem to be corrected
for absorption, even though this sample includes some (hot) Ap stars
as far away as 150 -- 180 pc, for which interstellar absorption is
probably significant. (However, for the five stars in common with
Lanz's sample, with distances in the range 90 -- 180 pc,
the average difference in integrated flux between the two studies is
only about 4\,\%.)  The three values of integrated flux reported by
Monier are not corrected for interstellar absorption, but all refer to
stars located between 50 and 100 pc from the Sun, and any
corrections should be at most a few percent.

\begin{figure}
\resizebox{9.0cm}{!}{
\includegraphics*[angle=270]{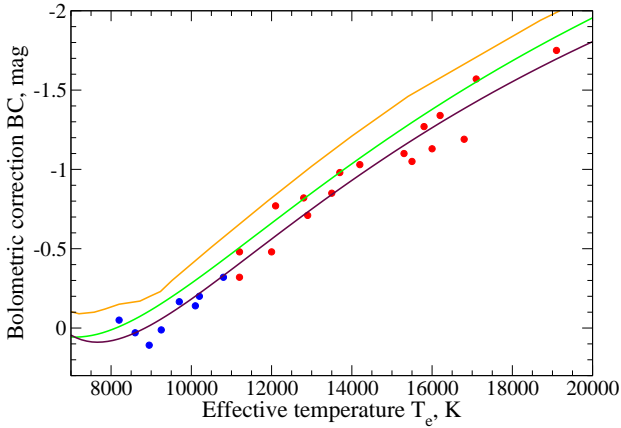}}
\caption{\label{bolcorr.fig} Bolometric corrections using integrated
  fluxes of magnetic Ap
  stars from Lanz (\cite{lanz84}, red circles), St\c{e}pei\'n \&
  Dominiczak (\cite{stepien89}, blue circles), and Monier
  (\cite{monier92}, blue circles) as functions of
  effective temperature. The upper solid line is the BC for normal
  stars from Schmidt-Kaler (Lang \cite{Lang92}); the middle line is
  Balona's (\cite{balona94}) fit to the fundamental data of Code et al.
  (\cite{code76}) and Malagnini et al. (\cite{malagnini86}), and the
  lowest curve is a polynomial fit to the Ap data as described in the
  text. }
\end{figure}

We have assigned effective temperatures to the stars for which
integrated fluxes are available using the same methods discussed
above, and computed the bolometric corrections for the stars of
St\c{e}pie\'n \& Dominiczak (\cite{stepien89}), and Monier
(\cite{monier92}) using the first equation in Lanz (\cite{lanz84}). 
Because of the lack of correction
for interstellar extinction, we have discarded all the stars from
St\c{e}pie\'n and Dominiczak's sample that have $\te >
11\,000$\,K.  However, we verified that including them did not
significantly change the results. 

Figure~\ref{bolcorr.fig} shows the resulting data set
of bolometric corrections as a function of \te\ together with 
our best fit (lower curve) to the Ap
star mean correction $BC_{\rm Ap}(\te)$:
\begin{equation}\label{bc-ap.eq}
   BC_{\rm Ap}(\te) = -4.891 + 15.147\, \theta 
      - 11.517\, \theta^2,
\end{equation}
where $\theta = 5040.0/\te$.

We also plot the 
variation of bolometric correction $BC_{\rm n}(\te)$ with \te\
for normal stars, using (upper curve) Schmidt-Kaler's values (Lang
\cite{Lang92}), and Balona's \cite{balona94} fit (middle curve) to the
fundamental data of Code et al. (\cite{code76}) and of Malagnini et al. 
(\cite{malagnini86}). 

Equation~(\ref{bc-ap.eq}) describes the available bolometric corrections for Ap
stars between about 7500 and 18000 K. The RMS deviation of the
individual Ap star data around this curve is about 0.1 mag,
considerably larger than that for normal stars around Balona's fit.
This reflects both larger uncertainty in both coordinates, and
probably also real and substantial star-to-star variations in the
bolometric correction for these rather diverse objects.

It is clear that the values of Code et al. (\cite{code76}) and
Malagnini et al. (\cite{malagnini86}) are the most accurate data for
the normal star bolometric correction $BC_{\rm n}(\te)$. Note
that the $BC$ values of Schmidt-Kaler, which are sometimes used for Ap
stars (e.g., Hubrig et al. \cite{Hubetal00}) are
systematically too negative compared to the more accurate data of Code
et al. (\cite{code76}) and Malagnini et al. (\cite{malagnini86}), and
differ from the calibration adopted here by about 0.25\,mag.

With typical distance modulus uncertainties of 0.2\,mag, an uncertainty
in the bolometric correction of about 0.1\,mag, and reddening
uncertainties of less than 0.1 mag, we estimate that the typical
uncertainty in $M_{\rm bol}$ is about 0.25\,mag, corresponding to an
uncertainty in \logl\ of about 0.1\,dex. We adopt this value as the
standard error of luminosity values in
Table~\ref{cl_stell_params_magmeas.tab}.

The apparent $V$ magnitudes and values of \te\ of stars in our
subsample have been used, together with the bolometric corrections of
Eq~(\ref{bc-ap.eq}), to determine the stellar luminosities \llo. For
most clusters we used the mean cluster reddening as tabulated by WEBDA
or one of the specific references. However, we found that using the
mean cluster or association reddening led in several cases to stars
appearing seriously underluminous compared to the position of the
expected main sequence. In these cases we have determined the
reddening $E(B-V)$ directly for each star from its position in a
colour-colour diagram, and used these values (with $A_V = 3.1 E(B-V)$)
to determine the absolute magnitude of each individual star.  This was
necessary for Ori OB1, the Upper Sco region of Sco OB2, and for NGC
2244, 2516, and 6193. In general, this led to a much tighter observed
main sequence.

\subsection{Magnetic field}

Magnetic fields of Ap stars may be determined through the measurement
of the Zeeman splitting of spectral lines in a simple intensity
spectrum, or through the analysis of circular polarization of spectral
lines.  The former method, exploited by Mathys et al.\
(\cite{Matetal97}), gives the average of the magnetic field modulus
over the visible stellar disk. Zeeman splitting may be detected only under
quite special circumstances, i.e., \vsini\ at most a few \kms, and
field strength at least 2\,kG, that are not met in most of the known
magnetic Ap stars. Circular polarization measurements are generally a
far more sensitive and broadly-applicable method of field detection,
and give the so called \textit{mean longitudinal magnetic field} \bz,
i.e., the component of the magnetic field along the line of sight
averaged over the stellar disk. 

Because of geometrical reasons, both mean field modulus and mean
longitudinal field change as the star rotates. In particular, a \bz\
measurement could be consistent with zero even in a star that
possesses a strong magnetic field. For the present work we need to
associate to each star a field value that is representative of the
stellar field strength.

For the few stars in our sample for which the full variation of the
longitudinal field \bz\ is known and has been fitted with a sine wave
of the form
\begin{equation}
   B_z(t) = B_0 + B_1 \sin(\omega t - \phi_0),
\end{equation}
we give the root-mean square longitudinal field strength $\brms
= (B_0^2 + B_1^2/2)^{1/2}$. For the great majority of the stars in the
table, however, only one measurement, or at most a few scattered
measurements, are available. In this case we simply report the raw
value of \brms\ averaged either over all available field
measurements, or over the subset with the smallest
uncertainties. Since most of the stars in this table have only one or
two measurements, this is clearly not a very accurate statistic for an
individual star, but averaged over our sample we believe that it
already provides a significant amount of useful information.

When the individual field strength values used in the computation of
\brms\ are taken from Paper I, the value derived from the full
spectrum is used if the measurement detects a field of less than 1~kG.
However, as discussed in Paper I, the metal lines in FORS1 spectra
yield an underestimate of the actual field strength for larger fields.
Accordingly, for fields above 1~kG we have used only the field
strength deduced from the Balmer lines.

Note that values of \brms\ smaller than about 250\,G typically should
be read as indicating that the RMS field estimated from the current
data set is of the order of 200\, G or less, regardless of the actual
tabulated value.

\section{Determining stellar masses and fractional ages}
\label{Sect_Age_Mass}

With effective temperatures and luminosities for almost all the stars
in Table~\ref{cl_stell_params_magmeas.tab}, we can now make comparison
with evolution tracks to determine approximate masses and fractional
ages (recall that the fractional age of a star is the fraction of its
main sequence lifetime, measured from the ZAMS to the TAMS, that has
already elapsed).

As discussed in Paper~I, there is a spread of chemical compositions
among nearby open clusters and associations of about a factor of 2.5
in metallicity, so that the accurate placement of a star on an
evolution track requires either the use of evolution tracks
appropriate to the particular star or stellar group, or consideration
of the additional uncertainties introduced by variation of evolution
tracks with metallicity. For studies of field stars, for which ages
are determined from comparison of position in the HR diagram with
evolution tracks, we showed in Paper~I that very important age
uncertainties arise from this effect, especially for stars in the
first half of their main sequence lives. We also showed that it is
only currently possible, with present uncertainties in \te\ and \llo,
to discriminate between field stars in the first and second halves of
their main sequence lifetimes. In contrast, uncertainty in bulk
chemical composition does not introduce an important additional
uncertainty into the determination of the stellar mass.

In this study we are determining stellar ages from the ages of the
clusters to which the stars of our sample belong. The principal
uncertainty for most of the available literature ages (0.1 --
0.2\,dex in $\log t$, as discussed above) arises from uncertainty as
to where to place the cluster turnoff in the HR diagram, or in the
appropriate models for pre-main sequence stars, rather than from
uncertainty in the composition that should be adopted for the
evolution tracks used for age determination. Consequently, we do not
need to use composition-specific evolution tracks for mass and
fractional age determination; we simply include the small uncertainty
carried by this effect in our overall uncertainty estimates. 

\begin{figure}
\scalebox{0.34}{
\rotatebox{270}{
\includegraphics*[3cm,0.1cm][20.7cm,26cm]{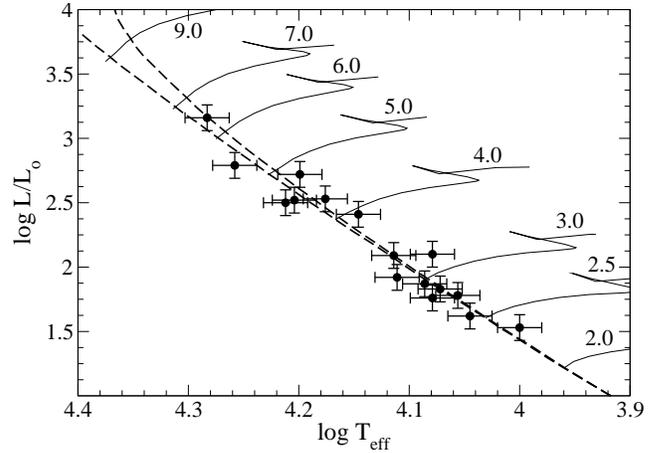}}}
\caption{\label{Cen-HRD-c-iso.fig} Comparison of computed positions of
  stars in the Lower Centaurus-Lupus, Upper Centaurus-Crux, and
  Upper Sco regions of the Sco OB2 association with evolution tracks
  for stars of $Z = 0.02$ and $M = 2, 2.5, 3, 4, 5, 6, 7,$ and 9
  $M_\odot$, and with bracketting isochrones for log ages of 6.9 and
  7.2, appropriate for these regions of the association.  }
\end{figure}

We then proceed as follows. Each star is placed in the HR diagram,
using the values of \te\ and \llo\ determined previously. Around each
star we construct an approximate error oval, using the adopted
uncertainties discussed above. The star positions are compared
graphically, one cluster at a time, with evolution tracks for stellar
masses in the range 1.7 to 9.0 $M/M_\odot$, and with the range of
isochrones allowed by the adopted cluster or association age $\pm 0.1$
 or 0.2\,dex, using models computed by the Padova group (see Fagotto et al.
\cite{fagetal94} and references therein) for $Z = 0.008$. 0.02 and
0.05. (These data are available in machine-readable form from CDS.)
This comparison allows us to test the hypothesis that each star is
actually a cluster member, and if it is, to determine its mass.

In almost all cases, the best overall agreement of individual star
positions in the HR diagram with computed isochrones is found for $Z =
0.02$. This agreement persists even when the magnetic Ap stars are
near the end of their main sequence evolution. This $Z$ value is
somewhat higher than the currently accepted best estimate for the Sun,
which is about $Z=0.012$ (Asplund et al.\ \cite{AspGreetal05}). It is
not clear whether our result indicates that $Z \approx 0.02$ is the
most appropriate mean value for young stellar groups in the solar
neighborhood, or whether this reflects a residual small systematic
error in our transformation to the theoretical HR diagram (for
example, modest underestimation of \te\ values as discussed above).
Note that the significantly better fit of the $Z = 0.02$ models to our
\te, \llo\ data compared to the fit with $Z = 0.008$ models indicates
that Padova models with $Z = 0.012$ would not fit our observational
data as well as the $Z = 0.02$ models do.  However, we cannot really
test how well our data would fit isochrones computed with the new
Asplund et al abundances, as these abundances have different ratios of
light (CNO) elements to iron peak elements than are assumed in any
available grid of evolutionary models.

We have tested the correctness of our computations, and assessed
whether our assumed uncertainties are realistic, by looking at the
results of such comparisons for nearby clusters and associations of
well-determined membership, precisely known distance and moderate or
small reddening. One such comparison is shown in
Figure~\ref{Cen-HRD-c-iso.fig} for the Sco OB2 association, using
tracks for $Z = 0.02$. This figure confirms the general correctness of
the transformations used in going from photometric data to the
theoretical HR diagram, and shows that our assumed uncertainties give
a reasonable description of the scatter about the $Z = 0.02$
isochrones.

In a few individual cases, typically in relatively distant or heavily
reddened clusters or associations, we find significant disagreement
between the deduced position of a star and the relevant isochrone
which strongly suggest that the star should be disqualified from
cluster membership on photometric grounds.  There are eight such cases
in Ori OB1 and five in various clusters.  All have been noted with
``?'' in the ``phot'' column of Table~\ref{cl_memb_pec_test.tab}, and
are omitted from Table~\ref{cl_stell_params_magmeas.tab}. The other
stars whose membership we have tested in this way, and whose positions
are consistent with the isochrones to within better than $2 \sigma$
are noted with ``y'' or ``p'' in this column. (We did not test stars
whose cluster membership or Ap nature was rejected on other grounds.)

Placing the observed stars on the appropriate range of cluster
isochrones allows us to provide a mass estimate for each. These masses
have been derived using the $Z = 0.02$ evolution tracks. Uncertainties
for these values, also reported in
Table~\ref{cl_stell_params_magmeas.tab}, arise primarily from the size
of the error oval of each stellar position due to uncertainties in
\te\ and \llo. A small contribution to the mass uncertainties is made
by the uncertainty in the $Z$ value appropriate for each cluster. For
stars whose $1 \sigma$ error ovals intersect the band of isochrones
defined by the (inexact) age of the cluster, we find that the
uncertainty in mass is typically about 5\,\% of the actual mass value.

Our method of mass determination may be clarified using
Figure~\ref{Cen-HRD-c-iso.fig}. Consider the star between the \mmo\ =
6 and 7 evolution tracks (\object{HD 125823}). It may be seen directly
from the figure that with this set of tracks and the indicated
uncertainties, the mass is between about 6.1 and 6.7 $M_\odot$, or
$\mmo = 6.4 \pm 0.3$. If the error oval of the star is a little above
or below the pair of isochrones for the age range of the cluster, the
mass uncertainty is increased from about 5\% to about 7\%.

As noted in Sect. 3.1, there are several clusters for which we have
determined a more precise age by forcing the cluster isochrone to pass
through the error oval of an Ap cluster member near the TAMS. The
method is very similar to that illustrated by Kochukhov \& Bagnulo
(\cite{KocBag06}) in their Figure~3. In these cases, we have included
in the estimate of cluster age uncertainty a contribution from the
uncertainty of the appropriate $Z$ value to use for the isochrones.
Even including this effect, in a few cases uncertainties in $\log t$
as small as $\pm 0.05$ -- 0.08 dex are obtained.

We then use the mass and its uncertainty to derive the expected main
sequence lifetime of each star, again using the $Z=0.02$ models.
Dividing the adopted cluster age by this lifetime, we obtain an
estimate of the fraction of the main sequence lifetime already elapsed
(the fractional age $\tau$) for each star. 

\begin{figure}
\includegraphics*[2.0cm,6.7cm][11.5cm,22cm]{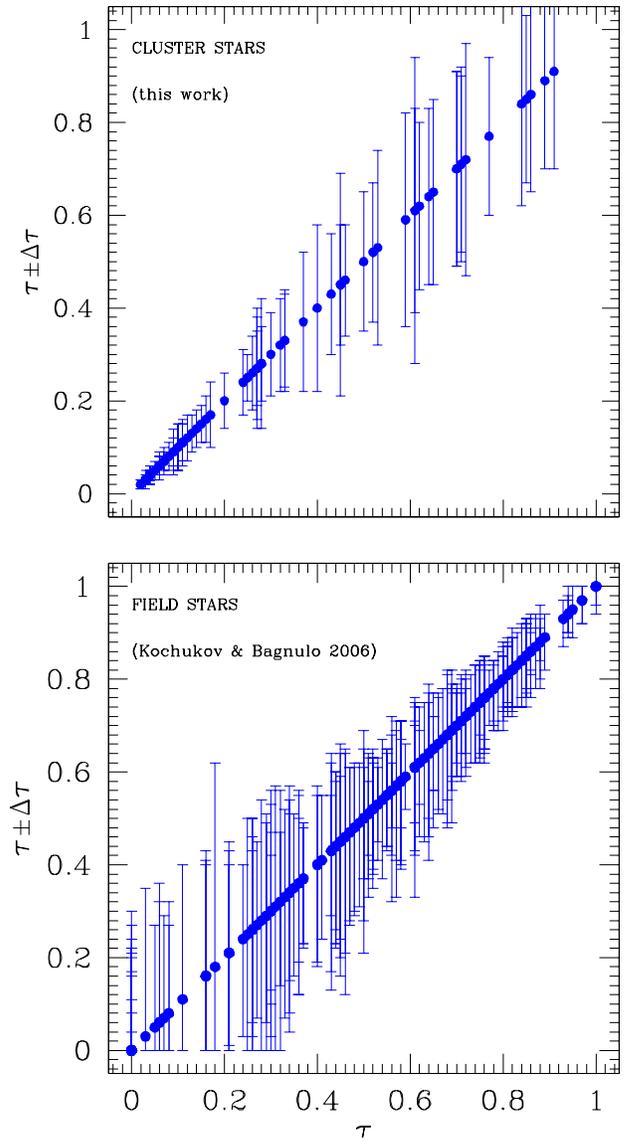}
\caption{\label{Fract-Err.fig} Upper panel: Distributions of the
  1\,$\sigma$ confidence interval of fractional age $\tau$ for our sample of
  open cluster stars. For comparison, we show also the same kind of
  distribution for the sample of field stars analysed by Kochukhov
  \& Bagnulo (\cite{KocBag06}). Note that Kochukhov \& Bagnulo assume
  considerably smaller errors in \te\ than we do, leading to smaller
  uncertainties in $\tau$, as discussed in the text.  }
\end{figure}

Several sources of error contribute to the uncertainty in this
quantity. Mass uncertainty of about 5\,\% leads to main sequence
lifetime uncertainty of about 12\,\% for evolution tracks for a given
$Z$. The uncertainty in the actual value of $Z$ for each cluster
introduces an additional lifetime uncertainty of up to about 7\,\%,
leading to a total uncertainty in main sequence lifetime of about $\pm
15$\,\%. Uncertainty in the ages of the clusters and associations has
been estimated in most cases to be approximately $\pm 0.1 - 0.2$\,dex,
or about $\pm 25 - 50$\,\%. In the clusters with age uncertainties in
this range, cluster age uncertainty dominates the total error of
$\tau$. In those clusters with one or more stars near the TAMS, for
which we can obtain more precise age estimates, the contributions from
cluster age uncertainty and from main sequence lifetime uncertainty
can be comparable.

We treat the uncertainties in the stellar lifetimes and the cluster
ages as independent.  For stars which are young relative to their main
sequence lifetimes ($\tau$ less than about 0.3) the accuracy of the
fractional age is substantially better than is possible at present by
placing field stars in the HR diagram, even for clusters with rather
poorly known ages. For stars of fractional age below about 0.1, the
improvement compared to the accuracy that can be obtained for field
stars is an order of magnitude or more. For $\tau$ larger than about
0.5 (these are the situations in which significant improvement in
cluster age precision, by isochrone fitting of individual evolved
stars, is possible), the uncertainty in fractional age of a cluster Ap
(for given error bars on \te\ and \llo, {\bf and comparable treatment
  of the $Z$ uncertainty}) is similar to that for nearby field stars,
since effectively in this circumstance the cluster age is most
precisely determined in the same way that the age of a field star is
determined.

The age confidence limits which characterise our sample are shown in
Fig.~\ref{Fract-Err.fig}, together with those obtained in the study
based on field stars by Kochukhov \& Bagnulo (\cite{KocBag06}).  It is
important to keep in mind when comparing these two distributions two
substantial differences between their analysis and ours. First, they
have assumed uncertainties for \te\ that are typically about 0.012 to
0.014 dex, compared to our uncertainty of 0.020 dex (they do not
include possible effects due to the overall uncertainty of the \te\
scale for Ap stars, which we consider an important probable source of
uncertainty). Increasing their \te\ uncertainty by 50\% would increase
the age uncertainty for evolved stars by a similar amount. Secondly,
they have not included any contribution from uncertainty in the bulk
$Z$ values of the stars of their sample (and thus uncertainty in which
evolution tracks to use for age determination), an effect which we
have found in Paper I to make an important contribution to the error
budget of ages for field stars. If Kochukhov \& Bagnulo had used the
same uncertainty estimates as we have, their $1\, \sigma$ confidence
limits would be very roughly a factor of two larger than those
shown in Fig.~\ref{Fract-Err.fig}, for all values of $\tau$.

In spite of these important differences, it is easily seen that the
precision of our fractional ages is considerably better than that of
Kochukhov \& Bagnulo for fractional ages below about 0.5.  For
fractional ages below about 0.2, the improvement is about an order of
magnitude, or more. Above $\tau \approx 0.5$, the fact that our
fractional ages are less precise than theirs is largely due to the two
differences in analysis noted above, since stellar main sequence
lifetimes and cluster ages for the most evolved stars are determined
in the same way as field star lifetimes and ages.

The stars of the final subsample are listed in
Table~\ref{cl_stell_params_magmeas.tab}, together with the data
derived for each star as discussed above. The first three columns of
this table characterise the clusters whose stars are in our subsample,
giving the cluster or association name, our adopted (logarithmic)
cluster age $\log t$ with uncertainty, and the true distance modulus
$DM$ of the cluster. The remaining columns list the stars which we
consider to be members with $\log \te$, \logl, \mmo\ and uncertainty,
fractional age $\tau$ and uncertainty, and the RMS estimate of the
longitudinal field \brms.

\section{Characteristics of our sample}\label{Sect_Character}

\begin{figure*}
\includegraphics*[2.3cm,16.5cm][26cm,22.8cm]{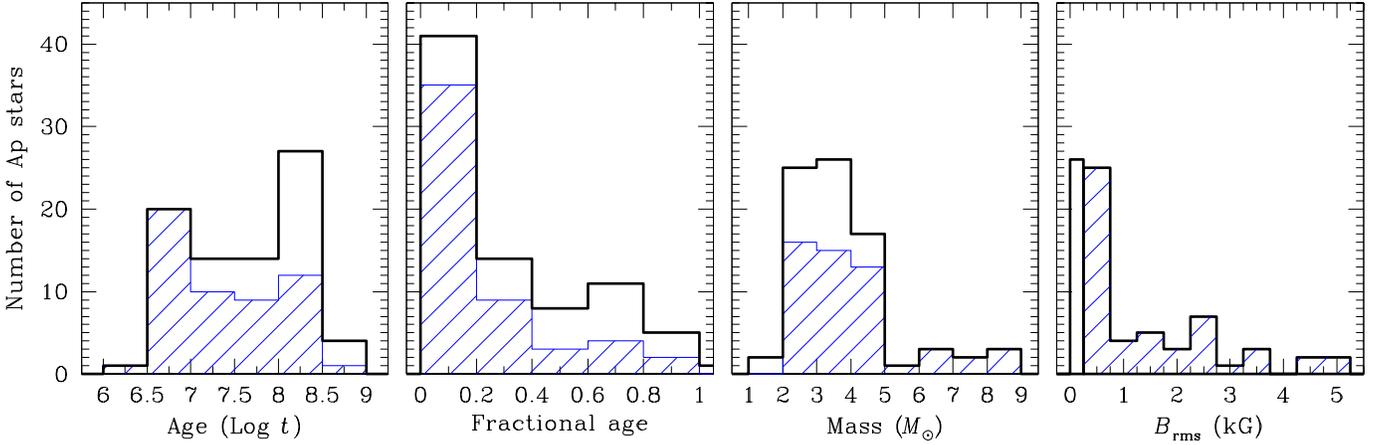}
\caption{\label{histos.fig} Distributions in our sample of age,
  fractional age, mass, and \brms.
In each box, the shaded histogram shows the distribution of those Ap
stars of Table~\ref{cl_stell_params_magmeas.tab} for which a
$B_{\rm rms} \ge 0.25$\,kG has been measured.
The thicker lines show the distribution of the entire sample.
}
\end{figure*}

It is of interest to note some distinctive properties of the data set
of Table~\ref{cl_stell_params_magmeas.tab} concerning age and mass
distributions.

\subsection{Absolute and fractional age distribution}
Because open clusters and especially associations tend to be disrupted
by galactic tidal acceleration on a time scale short compared to the
main sequence lifetime of an A star, we have a predominance of
relatively young stars (fractional ages less than 0.5) in our survey,
although the sample does include a significant number of more evolved
main sequence stars (especially at higher masses where the lifetimes
are shorter).  In fact, about half the stars in the sample have
fractional ages of less than 0.20. However, our data set is fairly
uniformly distributed in (log) absolute age between $\log t \approx
6.5$ and 8.5.  The distribution of absolute ages and of fractional
ages in our sample are shown in the first and second boxes on the left
of Fig.~\ref{histos.fig}, respectively.

\subsection{Mass distribution}
Because the distribution of masses in a cluster or association follows
the initial mass function (IMF), at least up to the cluster turn-off,
rather than the field mass function, which is more heavily weighted
towards low-mass stars of long main sequence lifetimes, our sample has
a relatively large number of hot, high-mass magnetic Bp stars. The
median mass in Table~\ref{cl_stell_params_magmeas.tab} is about
$3.5\,M_\odot$, and the median value of \te\ is about 12\,700 K.
These are stars that are not particularly common among the Ap-Bp stars
near the Sun, and may well prove to be rather interesting objects to
study individually. The distribution in mass of stars in our subsample
is shown in the third box in Fig.~\ref{histos.fig}. Notice that we
have only three stars (\object{CPD -32 13119}, \object{CPD -20 1640},
and \object{HD 66318}) that have $\mmo \leq 2.1$, and of these three,
only \object{HD 66318} has a detected (but huge) magnetic field (cf
Bagnulo, Landstreet, Lo Curto et al. \cite{Bagetal03}). Significant
numbers of Ap stars with detected magnetic fields begin to be found in
our sample only above about $2.3\, M_\odot$.

\subsection{Magnetic field distribution}
The distribution of magnetic field strengths found for probable
cluster Ap stars in our survey (including Ap stars in which no field
has yet been detected) is shown in the right box in
Fig.~\ref{histos.fig}. This distribution is qualitatively similar to
what is found for samples of field Ap stars (e.g. Bohlender \&
Landstreet \cite{BohLan90}), but scaled up to somewhat larger fields
than usual. The bulk of the stars (about 2/3 of the present sample)
have RMS fields of less than 1 kG, and the median field is about 450
G. This may be compared to the distribution of Bohlender \&
Landstreet; for a sample of 12 bright Ap field stars (also including
some in which fields had not yet been detected) they found none with
\brms\ above 1 kG, and a median field of about 300 G. The largest
difference between this sample and that of Bohlender \& Landstreet is
the presence of a substantial tail of high-field stars; almost a
quarter of the present sample have RMS fields of 2 kG or more.  The
differences between these two samples is probably due to the much
larger fraction of young Ap stars in the present sample (see the
discussion below).

\section{Discussion}
\label{Sect_Field_Evol}
\begin{figure*}
\scalebox{0.80}{
\includegraphics*[0.6cm,3.0cm][29.5cm,26.5cm]{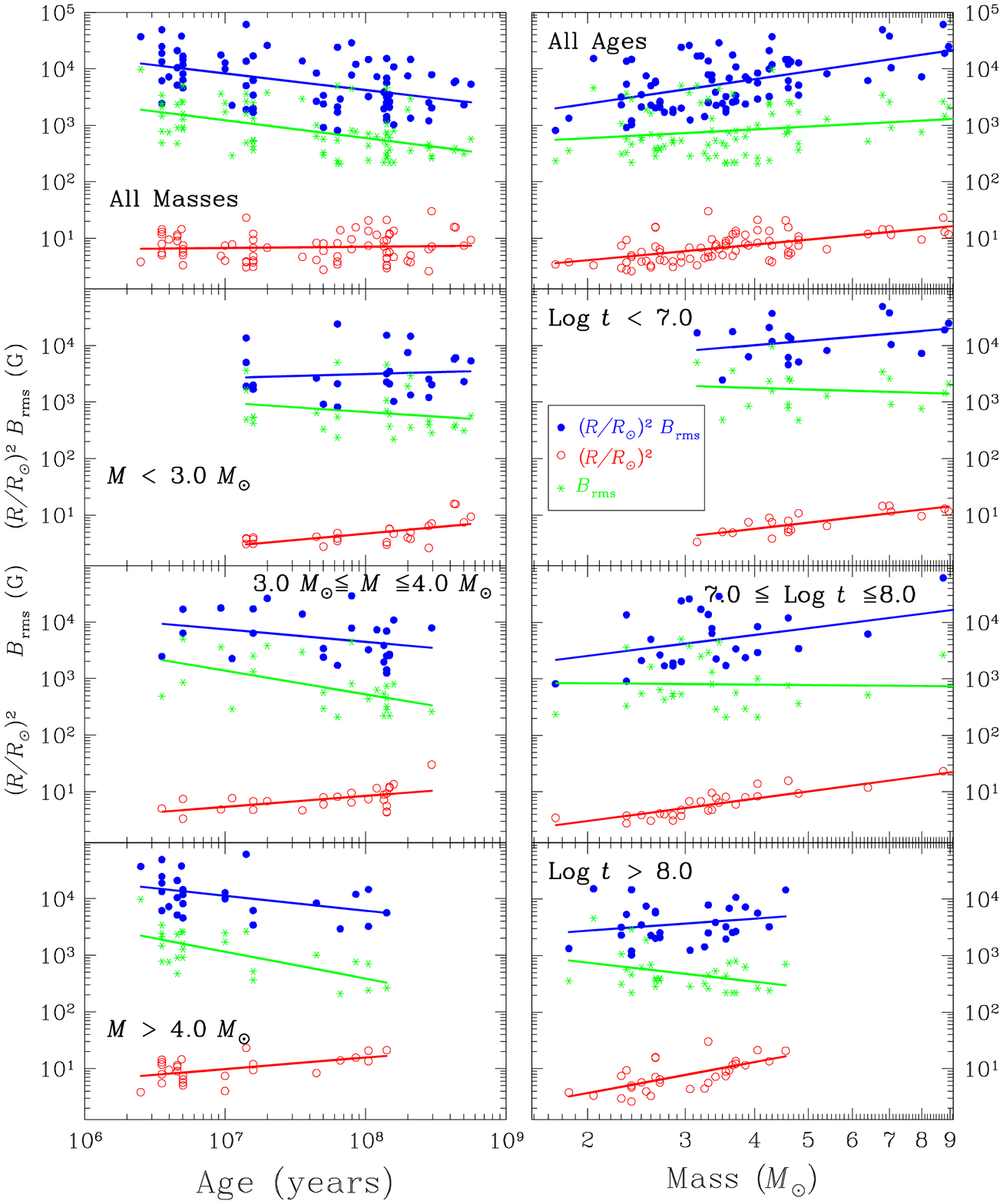}}
\caption{\label{regress_age_mass.fig} Plots of \brms\ (stars),
  $(R/R_\odot)^2$ (open circles), and $\brms (R/R_\odot)^2$
  (filled circles) are shown as functions of stellar
  age (left column) and mass (right column) for the subsample of
  Table~\ref{cl_stell_params_magmeas.tab}.  In each column the top
  panel shows all stars in the sample (except for $\theta^1$ Ori C).
  Lower panels show various cuts through the subsample as indicated on
  each panel. As discussed in the text, 200~G has been added to each
  value of \brms, in order to avoid having correlations made
  artificially steeper by very small values of \brms. 
}
\end{figure*}

We finally consider what general conclusions may be drawn from our
data about the evolution of magnetic fields up to and through the main
sequence phase of stellar evolution. With a sample of more than 80 stars
which are probable or definite cluster or association members,
probable or definite Ap-Bp stars, and for which at least one magnetic
measurement with $\sigma$ of the order of $10^2$ G is available, we
are now in a very good position to carry out a preliminary statistical
assessment of the evolution of Ap-Bp star magnetic fields based on
well-determined stellar ages.

\begin{table}
\caption{\label{log-log-fit-slope.tab} Slopes of regression lines in Fig.~\ref{regress_age_mass.fig}}
\begin{center}
\begin{tabular}{lllr@{\,$\pm$}l}
\hline
\hline
Sample     & $y$-axis              &  $x$-axis  &  \multicolumn{2}{c}{slope}\\
\hline
           & $\log (R^2\, B_{\rm rms})$ & $\log t$ &  $ -0.29 $&$ 0.07$ \\
All Masses & $\log R^2$                 & $\log t$ &  $  0.02 $&$ 0.04$ \\
           & $\log B_{\rm rms}$         & $\log t$ &  $ -0.31 $&$ 0.06$ \\ [2mm]

                   & $\log (R^2\,B_{\rm rms})$ & $\log t$ & $  0.07 $&$ 0.15$ \\
 $M < 3.0 M_\odot$ & $\log R^2$                & $\log t$ & $  0.23 $&$ 0.06$ \\
                   & $\log B_{\rm rms}$        & $\log t$ & $ -0.16 $&$ 0.14$ \\ [2mm]

                                     & $\log (R^2\,B_{\rm rms})$ & $\log t$ & $ -0.22 $&$ 0.14$ \\
 $3.0 M_\odot \le M \le 4.0 M_\odot$ & $\log R^2$                & $\log t$ & $  0.19 $&$ 0.06$ \\
                                     & $\log B_{\rm rms}$        & $\log t$ & $ -0.42 $&$ 0.14$ \\ [2mm]

                   & $\log (R^2\,B_{\rm rms})$ & $\log t$ & $ -0.27 $&$ 0.12$ \\
 $M > 4.0 M_\odot$ & $\log R^2$                & $\log t$ & $  0.21 $&$ 0.07$ \\
                   & $\log B_{\rm rms}$        & $\log t$ & $ -0.47 $&$ 0.11$ \\ [5mm]

          & $\log (R^2\,B_{\rm rms})$ & $\log (M/M_\odot)$ & $ 1.44 $&$ 0.26$ \\
 All Ages & $\log R^2$                & $\log (M/M_\odot)$ & $ 0.92 $&$ 0.14$ \\
          & $\log B_{\rm rms}$        & $\log (M/M_\odot)$ & $ 0.52 $&$ 0.29$ \\ [2mm]

               &$\log (R^2\,B_{\rm rms})$ & $\log (M/M_\odot)$ &  $  0.84 $&$ 0.57$  \\
$\log t < 7.0$ &$\log R^2$                & $\log (M/M_\odot)$ &  $  1.12 $&$ 0.21$  \\
               &$\log B_{\rm rms}$        & $\log (M/M_\odot)$ &  $ -0.29 $&$ 0.61$  \\ [2mm]

                         & $\log (R^2\,B_{\rm rms})$ & $\log (M/M_\odot)$ &$ 1.23 $&$ 0.52$  \\
$7.0 \le \log t \le 8.0$ & $\log R^2$                & $\log (M/M_\odot)$ &$ 1.32 $&$ 0.14$  \\
                         & $\log B_{\rm rms}$        & $\log (M/M_\odot)$ &$-0.09 $&$ 0.51$  \\ [2mm]

                & $\log (R^2\,B_{\rm rms})$ & $\log (M/M_\odot)$ & $  0.70 $&$ 0.61$  \\
 $\log t > 8.0$ & $\log R^2$                & $\log (M/M_\odot)$ & $  1.83 $&$ 0.38$  \\
                & $\log B_{\rm rms}$        & $\log (M/M_\odot)$ & $ -1.13 $&$ 0.57$  \\
\hline

\end{tabular}
\end{center}

\end{table}

The fundamental results of our survey are displayed in the eight
panels of Fig.~\ref{regress_age_mass.fig}, where we plot, using
logarithmic scales on all axes, the field strength \brms\ as a
function of stellar age (left column of the Figure), and as a function
of stellar mass (right column). In both columns, the top panel shows
all stars of the sample (except $\theta^1$ Ori C), while below we see
the same plot for three separate mass ranges (left column) or age
ranges (right column), with parameter ranges shown in each panel. 

The use of logarithmic scales would introduce a significant bias into
fits to these data, in that an RMS field which
happens to have a value of only a few~G is substantially lower than
the point for a star with an RMS field of, say 150\,G, although with
the uncertainties of our data, there is no statistical difference
between the two values. In order to avoid having small RMS field
values (essentially at present, non-detections) have an undue
influence on fits, we have artificially offset all \brms\ values by $+
200$\,G. 

\subsection{Does magnetic fields strength change with with time?}
The fundamental effect we are looking for with this study is to
discover whether typical RMS fields of magnetic Ap stars change with
stellar age, and if so, how they change. The top left panel of
Fig.~\ref{regress_age_mass.fig} appears to provide a first tentative
answer to this question. The variation of $\log \brms$ as a function
of $\log t$ has been fit with a linear function. This fit to the
(modified) data, shown as a line through the data points, was
found to have a slope which is non-zero (negative) at about the $5
\sigma$ level (see Table~\ref{log-log-fit-slope.tab}).  It appears
that the typical field strength decreases markedly with stellar age,
from a value of more than 1000\,G for stars having ages of about
$3\,10^6$~yr to a value of the order of 200\,G at $6\,10^8$~yr.

However, from Table~\ref{cl_stell_params_magmeas.tab}, it is clear
that most of the stars with very young ages (say, less than
$3\,10^7$~yr) are relatively massive (masses around 4 or 5~\mo ),
while the stars with ages of more than $10^8$ have smaller masses
(below about 4\,\mo).  Another symptom of this systematic change in
mass with age is seen in the variation of radii (in fact, we have
plotted the value of $(R/R_\odot)^2$) of the stars: the radius of a
single star increases with age, but the radii of the sample show no
significant variation with age, because the increase in radius of
individual stars is essentially compensated by the fact that older
stars have smaller masses and radii.

Thus we have considered three smaller sub-samples having limited mass
ranges: $M<3\,\mo$ (shown in the second panel from the top on the
left), $3\,\mo \le M \le 4\,\mo$ (third panel from the top), and $M >
4\,\mo$ (left bottom panel). Note that the subsample of stars defined
by $M<3\,\mo$ is in fact mainly populated with stars between 2.3 and
3\,\mo, and the subsample defined by $M>4,\mo$ in fact contains mainly
stars between 4 and 5\,\mo. The selected mass ranges are narrow enough
that the inhomogeneity of mass is not a major factor.

It is seen that for the two samples with larger masses (3 -- 4\,\mo\
and $> 4~\mo$) the \brms\ decreases with time with a slope that is
non-zero at the $3\,\sigma$ level. For the 3 -- 4~\mo\
sample, more than half of the stars younger than $4\,10^7$~yr have RMS
fields greater than 1\,kG, while all but one of the stars in this range older
than this age have fields below 1~kG. Similarly, in the most massive
sample, a majority of the stars with ages below $3\,10^7$ yr have
fields above 1\,kG while none of the older stars (with a range of ages
up to more than $10^8$~yr) have fields above 1\,kG.

In contrast, the lowest mass sample (mainly populated with stars
between 2.2 and 3\,\mo) does not show a significant decrease of RMS
field with age, and in fact there are about as many fields above 1\,kG
among the older (age $> 10^8$\,yr) stars as among the younger stars of
this mass range. Note that this group has no representatives among the
youngest clusters; we have not found any magnetic Ap stars with $M < 3
\mo$ in stellar groups of ages less than $10^7$~yr, although the Sco
OB2 association is close enough for such stars to be readily
identified.

One possible simple scenario for the observed decline of \brms\ with
age among stars of $M > 3 \mo$ would be that such stars
(approximately) conserve magnetic flux as they evolve through the main
sequence phase. During the evolution from ZAMS to TAMS, the radius of
a star increases by a factor of roughly 3. 

We have tested our samples to see if flux $\Phi_{\rm B} \sim \brms
(R/R_\odot)^2$ is statistically conserved as we look at stars of various ages.
In the various panels of Fig.~\ref{regress_age_mass.fig}, we plot both
$(R/R_\odot)^2$ and $\brms R^2$.  In the left column of panels, the individual
mass bands all show increasing values of $(R/R_\odot)^2$ with increasing age at
the $3 \sigma$ level, although the rise is not as great as expected,
probably due to the increasing contribution of lower mass stars in
each sample as we go to greater ages. In the two panels at the lower
left, where we find strongly significant decline in \brms, the value
of flux $\brms R^2$ appears to decline somewhat, but the decline is
only present at about the $2 \sigma$ level. We may regard this as consistent
either with some real decline in magnetic flux, or with conservation
of magnetic flux. 

To summarise the results of this analysis, we find that magnetic Ap
stars with masses above 3~\mo\ have fields which decline substantially
over an age of the order of $2 - 3\,10^7$~yr. For these stars, the
magnetic flux may decline with time, or remain roughly constant. The
RMS fields of less massive stars change little over time-scales
approaching $10^9$~yr, and there is no strong evidence that the
magnetic fluxes change significantly over this time scale.

The evolution with time of fields in main sequence Ap stars has also
been discussed by Kochukhov \& Bagnulo (\cite{KocBag06}). They find
that the value of \brms\ declines with increasing age for all their
subsamples of Ap stars except for those with $\mmo < 2$, in contrast
to our result that \brms\ decreases with age except for stars of $\mmo
< 3$. Thus the two studies agree that there is a significant decrease
of \brms\ with time among Ap stars at least above $3 M_\odot$, although
they disagree on precisely where  this behaviour changes over to more
nearly constant fields. 

However, Kochukhov \& Bagnulo (\cite{KocBag06}) find a significant
increase in magnetic flux with age for stars of $\mmo < 3$, while we
find, if anything, a modest decrease of flux with time. At this point,
it is probably fair to say that approximate magnetic flux conservation
during the main sequence phase is not ruled out. It will be very
useful to obtain more precise values of \brms\ for stars in the
cluster sample (by getting multiple observations of each star, and
expanding the sample of stars measured) to resolve the differences
between these two studies.

\subsubsection{Do magnetic fields appear late in main sequence evolution?}  
So far we have discussed a possible decline of the magnetic field, but
we have not commented on whether the magnetic field is present at the
very moment when a star reaches the ZAMS, or whether the field appears
at the stellar surface at some later stage in its main sequence
evolution. This problem has been previously discussed, among others,
by Hubrig et al.\ (\cite{Hubetal00}), Bagnulo et al
(\cite{Bagetal03}), P\"{o}hnl et al.~(\cite{PPM05}), and Kochukhov \&
Bagnulo (\cite{KocBag06}).

Based on the analysis of field stars, Hubrig et al.\
(\cite{Hubetal00}) have proposed that, in stars with $M \la
3\,M_\odot$, a magnetic field appears at the surface of an Ap stars
only after about 30\,\% of its main sequence liftime has elapsed. This
conclusion was later reaffirmed by Hubrig et al.~(\cite{HNS05};
\cite{HSN05}). However, this result was contradicted by the discovery
of a field in the young $2.1 M_\odot$ star \object{HD 66318} (Bagnulo et al.
\cite{Bagetal03}. Furthermore, both P\"{o}hnl et al.~(\cite{PPM05})
and Kochukhov \& Bagnulo (\cite{KocBag06}) were unable to reproduce
the results of Hubrig et al. of late emergence of surface fields, but
both studies found a remarkable shortage of young magnetic stars with
$M \la 2\,M_\odot$.

\begin{figure}
\scalebox{0.44}{
\includegraphics*[0.3cm,5cm][21cm,25cm]{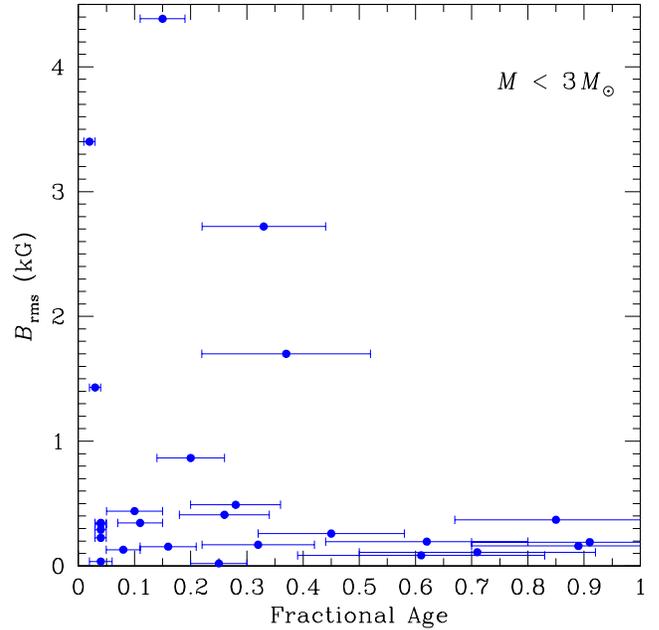}}
\caption{\label{frac3m.fig} The position of the stars with
$M \le 3\,M_\odot$ in the diagram of $B_{\rm rms}$ vs.\ fractional age.
}
\end{figure}

In the present sample we have a total of six stars with masses
definitely below $3 M_\odot$, fractional ages below 0.30, and
unambiguously detected magnetic fields. These stars are
\object{HD\,65712} (Paper I), \object{HD\,66318} (Bagnulo et al
\cite{Bagetal03}), \object{HD\,92385} {\bf (Paper I)},
\object{HD\,112381} {\bf (Bohlender et al \cite{Bohetal93})},
\object{HD\,119419} (Thompson, Brown \& Landstreet \cite{Thoetal87}),
and \object{HD\,318100} (Paper I). One additional star,
\object{HD\,74169} (Paper I), satisfies all criteria but only has one
measurement with field detection at the few sigma level, and still
ought to be confirmed.  Figure~\ref{frac3m.fig} shows the position of
all of the stars of our sample with $M \le 3\,M_\odot$ in the diagram
of $B_{\rm rms}$ vs.\ fractional age. Our study thus definitely
contradicts that of Hubrig et al. concerning late field emergence.

We have investigated the reasons for the discrepancies between our
conclusions and those of 
Hubrig et al.~(\cite{Hubetal00}) and we have identified several points
that may have led them to obtain incorrect results.

\textit{i)} Hubrig et al.(\cite{Hubetal00}) have used the bolometric
corrections of Schmidt-Kaler, which are too large even compared to the
most accurate values for normal stars, and lead to an overestimate of
bolometric magnitude relative to the BC's adopted here (see
Fig.~\ref{bolcorr.fig}) by about 0.25 mag (0.1\,dex in \logl). Using
more appropriate BC's would move their observed stars downwards,
towards the ZAMS, and younger ages. \\
\textit{ii)} As discussed above, the whole effective temperature scale
for magnetic Ap stars may be systematically a few percent low. This
possibility has little effect on ages derived for cluster stars, but,
if correct, would move the field stars towards the ZAMS as well. \\
\textit{iii)} Hubrig et al. (\cite{Hubetal00}) applied the Lutz-Kelker
correction to their data, moving the stars upwards in the HR diagram,
away from the ZAMS. As discussed by St\c{e}pie\'n (\cite{Ste04}) and
Kochukhov \& Bagnulo (\cite{KocBag06}), this correction should not
have been used. Removing it also moves the
Hubrig et al. sample closer to the ZAMS.  \\
\textit{iv)} Finally, in Hubrig et al.~(\cite{Hubetal00}) there are a
two significant errors: the \te\ value of \object{HD\,119419} is about
15\,\% too low, and that of \object{HD\,147010} is more than 50\,\% too small.
Correcting these errors would move \object{HD\,147010} to a mass slightly above
3$M_\odot$ but to a position close to the ZAMS. \object{HD\,119419} would
have remained well below $3\,M_\odot$ and moved close to the ZAMS,
providing Hubrig et al. with a counter-example to their hypothesis in
their own data.

The changes to the data of Hubrig et al.~(\cite{Hubetal00}) resulting
from the points discussed above appear to be enough to make the effect
they report disappear from their data. This result, together with the
results of Bagnulo et al. (\cite{Bagetal03}), P\"{o}hnl et
al.~(\cite{PPM05}), Kochukhov \& Bagnulo (\cite{KocBag06}), and our
own survey, lead us to reject the hypothesis that stars with $M \le
3\,M_\odot$ becomes magnetic only after they have spent a significant
fraction of their life in the main sequence.

In our survey we note a relatively small, but non-zero, number of
stars of mass below about $2 M_\odot$ ($\te < 8500$ K). Only three of
the 81 stars in Table~\ref{cl_stell_params_magmeas.tab} have masses in
the range $1.7 \leq \mmo \leq 2.1$. To some extent this may reflect
the absolute rarity of such stars; in the field, the Ap stars near F0
make up only about 1\,\% of all stars, compared to about 10\,\% near
A0 (Wolff \cite{Wol68}). Still, a fairly large number of such stars
are now known in the field, both as a result of searches for roAp
stars (Kurtz \& Martinez \cite{KurtzMartinez93}) and from searches for
magnetic stars showing visible Zeeman splitting, which seems to be
commoner among low-mass Ap stars (Hubrig et al. \cite{Hubetal00}). A
number of such low-mass Ap stars are also found in the survey of
Kochukhov \& Bagnulo (\cite{KocBag06}); the rarity of low-mass Ap
stars in our sample seems to be anomalous. It may result primarily
from selection effects. Such stars will be among the faintest Ap stars
in most of the clusters we have studied, and may simply not have been
selected as Ap stars in identification studies, either because they
were near the threshold of the study, or because the method used
(such as the $\Delta a$ method) is not very sensitive to the coolest
Ap stars.

Note that, in contrast to the results of P\"{o}hnl et
al.~(\cite{PPM05}) and Kochukhov \& Bagnulo (\cite{KocBag06}), both of
whom found a deficiency of {\em young} Ap stars with masses below $2
M_\odot$, all three of the low-mass stars in our survey are fairly
close to the ZAMS; the largest fractional age among these stars is
0.16. This effect probably results from the fact that the main
sequence lifetime of such stars ($10^9$ yr or more) is considerably
longer than the ages of most clusters in our sample. Another
distinctive feature of our sample of particularly low-mass Ap stars is
that no field has so far been definitely detected in either of the two
stars with masses below $2 M_\odot$. The lowest mass cluster Ap star
for which a field has been definitely detected is \object{HD 66318}
(Bagnulo et al. \cite{Bagetal03}). This is quite different from the
situation in the field, where a number of Ap stars with detected
fields have masses that extend down to about $1.5 M_\odot$ (cf
Kochukhov \& Bagnulo).

From the present study (in particular from the fractional ages
reported in Table~\ref{cl_stell_params_magmeas.tab}) it is clear that
magnetic fields are present at essentially the zero age main sequence
phase for all stellar masses between about 2 and $5 M_\odot$; stars
with fractional ages below about 0.05 occur throughout this range. It
seems very reasonable to suppose that these fields are already present
late in the pre-main sequence phase, and this has recently been
confirmed by the discovery of ordered global fields in several
pre-main sequence Herbig AeBe stars (Donati et al. \cite{DonSem97};
Wade et al. \cite{WadDro05}; Catala et al. \cite{CatAle06}).

\subsection{Do magnetic fields depend on mass?}
A further issue that we may address with our data is the question of
whether there is a variation with mass of typical RMS field strength,
at a given stellar age.  Thompson, Brown \& Landstreet
(\cite{Thoetal87}) have argued that they detected such an effect, with
stars more massive than roughly 6 or 7\,\mo\ having fields about a
factor of two larger than lower mass stars. Kochukhov \& Bagnulo
(\cite{KocBag06}) find that the field strength \brms\ shows no clear
variation with mass, but argue that the magnetic flux definitely
increases with mass.

A plot of the RMS fields of all the stars in our sample as a function
of stellar mass (top right panel in Fig.~\ref{regress_age_mass.fig})
appears to show a trend of increasing field strength with
mass. However, again we must recall that this sample has considerable
change in typical mass as a function of age. To reduce the influence
of this effect, the whole sample is subdivided into three samples of
more limited age range: $\log t < 7.0$, $7.0 \le \log t \le 8.0$, and
$\log t > 8.0$. As noted above,
the youngest sample contains only stars more massive than about 3~\mo,
and the oldest sample is restricted to stars less massive than
4~\mo. None of these smaller samples of restricted age shows a clearly
significant trend of RMS field with mass. In particular, we have six
stars in the youngest sample with masses above the value at which
Thompson et al. (\cite{Thoetal87}) thought they had found a jump in
field strength, but we see no significant change at this mass.

From our data, it is not clear if magnetic flux is typically a
function of stellar mass. The age range between $\log t = 7$ and 8
appears to show a significant increase of magnetic flux with
mass, but this does not seem to be present (over more restricted  mass
ranges) for younger or older samples.

\section{Conclusions}
We have assembled a sample of nearly 160 stars for which magnetic
field measurements are available (primarily from Paper I) and which
are {\em possible} members of clusters or associations, and {\em
  possible} magnetic Ap or Bp stars. After detailed examination of
astrometric, photometric, and spectroscopic data available for these
stars, 81 remain as {\em probable} cluster members and {\em probable}
magnetic Ap stars. These stars are listed in
Table~\ref{cl_stell_params_magmeas.tab}.

For these stars, we have obtained cluster/association distances and
ages (typically accurate to $\pm 0.05 - 0.2$\,dex) from the literature
and from our own isochrone fits to stars near the TAMS. We have
determined effective temperatures (typically accurate to $\pm
0.02$\,dex) for almost all of these stars from both Str\"omgren and
Geneva photometry. We have derived new bolometric corrections for Ap
stars, and obtained luminosities of individual stars with typical
uncertainties of $\pm 0.1$\,dex.

By placing the stars of this subsample in the HR diagrams of their
clusters, we have determined their masses with an accuracy of about
$\pm 5$\,\%, and improved the precision of the ages of several
clusters. The mass values allow us to obtain main sequence lifetimes
from evolutionary calculations, with an overall uncertainty (including
the uncertainty in bulk chemical composition) of about $\pm 15$\,\%.
From main sequence lifetimes and cluster ages, we deduce fractional
ages $\tau$ (the fraction of the main sequence lifetime of each star
already elapsed). The uncertainty in the fractional age is almost
always less than half of the actual fractional age, and thus for stars
near the beginning of their main sequence lives, the fractional ages
are {\em much} more accurate than ages than can be derived for field
stars by placing them in the HR diagram (see the penultimate columns
of Table~\ref{cl_stell_params_magmeas.tab}).

From the stellar characteristics derived here we have assembled for
the first time a substantial sample of magnetic Ap stars for which the
absolute and fractional ages are known with really useful precision.
These accurate masses and fractional ages allow us to definitively test
the hypothesis of Hubrig et al. (\cite{Hubetal00}) that in Ap stars of
masses $M < 3 \mo$, the magnetic fields first become visible at the
surface after about 30\,\% of the main sequence lifetime has elapsed. We
have identified six clear counter-examples to this hypothesis in our
sample. As a result, we are able to definitively rule out the hypothesis
of late field emergence in low-mass Ap stars.

Although the field measurements of the stars of our sample are still
rather incomplete, we can use the available data to constrain the
typical evolution of magnetic field strength with stellar age. We find
that fields are definitely present essentially at the ZAMS (fractional
ages of 0.05 or less) for Ap stars ranging between 2 and 5~\mo, and
for fractional ages of less than about 0.10 for masses up to 9~\mo.

The evolution of magnetic fields with time suggested by our data
reveals a quite unexpected pattern. For stars of $M > 3 \mo$ we find
strong evidence that the field strength declines by a factor of a few
on a timescale of about $2 - 3\,10^7$ yr, in agreement with the
conclusion of Kochukhov \& Bagnulo (\cite{KocBag06}). It is not clear
for these stars whether the total magnetic flux remains approximately
constant, causing \brms\ to decrease as the stellar radius expands, or
whether this flux actually declines slowly.

In contrast, for stars of $M < 3 \mo$, there is no convincing evidence
in our data of field strength decrease, nor of magnetic flux decrease,
even on a time scale of several times $10^8$ yr. This conclusion, very
different from the result for more massive Ap stars, represents a
significant puzzle; perhaps this result reveals some important
structural difference between the lower and higher mass Ap stars. 

Finally, we are able to examine the question of whether there is a
significant variation in typical magnetic field strength as a function
of mass within a given age band. We do not find any strong evidence
for such a variation, in contrast to Thompson et al. (\cite{Thoetal87}),
who argued that typical field strength increases rather abruptly at
about 6 -- 7~\mo. 

These results indicate clearly that studies of magnetic stars in
clusters can yield much useful information about the nature and
evolution of their fields. Further observations, both to detect fields
in more cluster Ap stars (thereby increasing the size and diversity of
the sample), and to improve the accuracy of the RMS fields in stars
already detected, are already in progress.

\acknowledgements{
Work by JDL, JS, and GAW has been supported by the Natural Sciences 
and Engineering
Research Council of Canada. LF has received support from the Austrian 
Science Foundation
(FWF project P17980-N2). This work is based on data collected at
the ESO VLT within the context of programs 068.D-0403, 070.D-0352,
272.D-5026, 073.D-0498, and 074.D-0488, and on data retrieved from the
ESO/ST-ECF archive.  This research has made use of the SIMBAD
database, operated at CDS, Strasbourg, France.
}

\Online
%
\begin{longtable}[c]{ll|llcc|cccccccllll}
  \caption{\label{cl_memb_pec_test.tab} Tests of cluster membership
    and chemical peculiarity.  The first two columns give the cluster
    or association to which the star may belong, and a common
    designation of the star. Columns 3, 4 and 5 indicate whether the
    measured parallax, proper motions, {\bf and position in the HR
      diagram} are consistent with membership, and column 6 gives our
    conclusion about cluster membership (y = yes, p = probable, ? =
    questionable, n = no, blank = no data; letters in parentheses
    refer to references at end of table). Columns 7 through 10
    summarize evidence that each star is an Ap star according to its
    spectral classification, to the value of $\Delta a$ and/or $Z$, to
    the presence of periodic variability, and to whether a magnetic
    field is actually detected (letters in parentheses in the ``mag
    fld'' column refer to references at the end of the table). Column
    11 reports our conclusion as to whether the star is indeed an Ap
    star.
  } \\
  \hline \hline
  Cluster or         &   Star        & \multicolumn{4}{|c|}{Cluster membership} &        \multicolumn{5}{c}{Magnetic Ap star} \\
  Association        &               & $\pi$ & $\mu$ & phot &  member?      &  Sp      & $\Delta a - Z$ & var & mag fld & Ap?  \\
  \hline
\endfirsthead
\hline
Cluster or         &   Star        & \multicolumn{4}{|c|}{Cluster membership} &        \multicolumn{5}{c}{Magnetic Ap star} \\
Association        &               & $\pi$ & $\mu$ & phot &  member?      &  Sp      & $\Delta a\, ,\, Z$ & var & mag fld & Ap?  \\   
\hline     
\endhead
\hline
\endfoot
\hline
\hline
\endlastfoot
Blanco 1           &    HD 225264  & p    & y    &       & y (R, B, D, K) & ?        & n        & n        & n (F)    & ?  \\
$\alpha$ Per       &     HD 19805  & y    & p    & y     & y (Z, K)       & ?        & p        & n        & ? (a, r) & ?*    \\ 
                   &     HD 20135  &      & n    &       & n (K)          & p        & y        &          & n (a)    & p    \\
                   &     HD 21699  & y    & p    & y     & y (R, Z, K)    & y        & y        & y        & y (b)    & y    \\ 
                   &     HD 22401  & y    & p    &       & p (Z, K)       & p        & n        & n        & n (a)    & ?    \\ 
Pleiades           &    HD 23387A  &      & y    & p     & y (D, K)       & p        & ?        &          & ? (a, r) & ?*   \\ 
                   &     HD 23408  & y    & y    &       & y (D, K)       & ?        & n        & n        & n (a, s, F) & ?    \\ 
NGC 1662           &     HD 30598  & y    & y    & y     & y (D, K)       & p        & y        & n        & n (F)    & p    \\ 
Ori OB1a           &     HD 35008  & p    & y    & ?     & ? (Br)         & p        & n        & n        & y (F)    & y    \\ 
                   &     HD 35298  & y    & y    & p     & p (Br)         & y        & y        & y        & y (c)    & y    \\ 
                   &     HD 35456  & y    & y    & ?     & ? (Br)         & y        & y        & p        & y (c)    & y    \\ 
                   &     HD 35502  & y    & y    & y     & p (Br)         & p        & p        & p        & y (c)    & y    \\ 
                   &     HD 36429  & y    & y    &       & p (Br)         & p        & n        & n        & ? (c)    & ?    \\ 
                   &     HD 36549  & y    & p    &       & p (Br)         & y        & ?        & n        & n (F)    & ?    \\ 
Ori OB1b           &     HD 36046  & p    & y    &       & p (Br)         & y        & n        & n        & n (F)    & ?   \\ 
                   &     HD 36313  &      & y    & ?     & ? (Br)         & y        & y        & y        & y (c)    & y    \\ 
                   &     HD 36485  &      & p    & y     & p (Br)         & y        &          & y        & y (d)    & y    \\ 
                   &     HD 36526  &      & y    & y     & p (Br)         & y        & p        & y        & y (c)    & y    \\ 
                   &     HD 36668  & y    & y    & ?     & ? (Br)         & y        & y        & y        & y (c)    & y    \\ 
                   &    HD 290665  &      & y    & ?     & ? (Br)         & y        & y        &          & y (F)    & y    \\ 
                   &     HD 37140  &      & y    & y     & p (Br)         & y        & p        & y        & n (c)    & y    \\ 
                   &     HD 37333  & y    & y    &       & p (Br)         & n        & y        & n        & n (F)    & ?    \\ 
                   &     HD 37479  &      & y    & y     & p (Br)         & y        &          & y        & y (d)    & y    \\ 
                   &     HD 37633  &      & p    & y     & p (Br)         & y        & y        &          & y (F)    & y    \\ 
                   &     HD 37776  & y    & y    & y     & p (Br)         & y        &          & y        & y (g)    & y    \\ 
Ori OB1c           &     HD 36540  & y    & y    & p     & p (Br)         & y        & n        & y        & y (F)    & y    \\ 
                   &     HD 36629  & p    & y    &       & p (Br, T)      & p        &          & n        & n (F)    & ?    \\ 
                   &     HD 36918  &      & y    &       & p (Br, T)      & ?        & n        &          & n (F)    & ?    \\ 
                   &     HD 36916  & y    & p    & ?     & ? (Br)         & y        &          & y        & y (c, F) & y    \\ 
                   &     HD 36960  & y    & y    &       & p (Br, T)      & p        &          & n        & n (F)    & ?    \\ 
                   &     HD 37017  & y    & y    & ?     & ? (Br)         & y        &          & y        & y (d)    & y    \\ 
                   &     HD 37058  &      & y    & y     & p (Br)         & p        &          & y        & y (k, F) & y    \\ 
                   &     HD 37210  &      & y    & ?     & ? (Br)         & y        & y        & y        & n (c, F) & y    \\ 
                   &     HD 37470  & p    & y    &       & p (Br)         & ?        &          & n        & n (c, F) & ?    \\ 
                   &     HD 37642  & y    & y    & p     & p (Br)         & y        & y        & y        & y (c)    & y    \\ 
Ori OB1d           &     HD 36982  &      & y    &       & y (Br, T)      & p        &          &          & n (F)    & ?    \\ 
                   &     HD 37022  &      & y    & y     & y (Br, *)      & n        &          & y        & y (e, F) & y    \\ 
NGC 2169           &   NGC 2169 12 &      & p    & y     & p (K)          & p        & p        & y        & y (F)    & y    \\ 
NGC 2232           &     HD 45583  & y    & y    & y     & y (R, B, K)    & y        & y        & y        & y (F)    & y    \\ 
NGC 2244           &  NGC 2244 334 &      & y    & p     & y (M, S)       & y        & y        &          & y (F)    & y    \\ 
NGC 2287           &     HD 49023  & p    & y    &       & y (B, D, K)    & p        &          & n        & n (F)    & ?    \\ 
                   &  CpD-20 1640  &      & p    & y     & p (D, K, *)    & p        & n        &          & n (F)    & p    \\ 
                   &     HD 49299  &      & p    & y     & p (D, K)       & y        & y        &          & y (F)    & y    \\ 
                   &     HD 49333  & n    & n    &       & n (B, D, K)    & y        &          & y        & y (h, k) & y    \\ 
CMa OB2            &     HD 51088  &      & p    & p     & p (D, K)       & y        & y        &          & ? (F)    & y    \\ 
NGC 2323           &     HD 52965  &      & p    & y     & p (B, D, K)    & p        &          & n        & n (F)    & p    \\ 
Cr 132             &     HD 56343  & p    & y    & p     & p (B, *)       & n        & y        & n        & y (F)    & y    \\ 
Cr 135             &     HD 58260  & n    & n    &       & n (B, D, K)    & y        &          &          & y (d)    & y    \\ 
NGC 2422           &   BD-14 2015  &      & p    &       & p (K)          & p        & n        & n        & n (F)    & ?    \\ 
                   &     HD 61045  & y    & y    & y     & y (R, B, D, K) & y        & y        & n        & y (F)    & y    \\ 
                   &   BD-14 2040  &      & y    &       & y (K)          & p        & n        &          & n (F)    & n    \\ 
                   &   BD-14 2028  &      &      &       & p              & y        & n        & n        & n (F)    & ?    \\ 
NGC 2451A          &     HD 62376  & y    & n    &       & n (B, K, *)    & n        & y        & y        & n (F)    & p    \\ 
                   &     HD 63079  & y    & y    &       & y (B, K)       & p        & n        & n        & n (F)    & ?    \\ 
                   &     HD 63401  & y    & p    & y     & y (R, B, K)    & p        & y        & y        & y (s, F) & y    \\ 
NGC 2451B          &   CD-37 3845  &      & p    &       & p (C)          & p        & ?        &          & n (F)    & ?    \\ 
                   &     HD 62992  & p    & n    &       & n (C, K)       & y        & y        &          & ? (F)    & y    \\ 
NGC 2489           &  NGC 2489 58  &      &      &       & ?              & ?        & y        &          & n (F)    & ?    \\
                   &  NGC 2489 40  &      &      & y     & ?              & ?        & y        &          & n (F)    & ?    \\ 
NGC 2516           &     HD 65712  &      & y    & y     & y (K, *)       & y        & y        &          & y (F)    & y    \\ 
                   &   CpD-60 944A &      & y    & y     & y (D, K)       & y        & y        &          & p (F)    & y    \\ 
                   &   CpD-60 944B &      & p    & p     & p (D, K, *)    & p        & y        &          & n (F)    & y    \\ 
                   &     HD 65949  & p    & p    &       & p (D, K, *)    & n        & n        &          & n (F)    & n    \\
                   &   CpD-60 978  &      & p    & y     & p (D, K)       & y        & y        & y        & n (F)    & y    \\ 
                   &     HD 65987  &      & y    & y     & y (D, K)       & y        & y        & p        & y (F)    & y    \\ 
                   &   CpD-60 981  &      &      &       & ? (K)          & ?        & ?        &          & n (F)    & ?    \\ 
                   &     HD 66295  &      & y    & y     & y (D, K)       & y        & y        & y        & y (F)    & y    \\ 
                   &     HD 66318  &      & p    & y     & p (D, K)       & y        & y        & n        & y (F)    & y    \\ 
NGC 2546           &  NGC 2546 258 &      &      &       & ?              & ?        & y        &          & n (F)    & p    \\ 
                   &  NGC 2546 201 &      &      &       & p (K)          &          &          &          & y (F)    & y    \\ 
                   &  NGC 2546 197 &      & y    & y     & y (D, K)       & p        & y        &          & n (F)    & p    \\ 
                   &     HD 69004  &      & p    & ?     & ? (D, K)       & y        & y        &          & n (F)    & p    \\ 
                   &     HD 69067  &      & n    &       & n (D, K)       & y        & y        &          & y (F)    & y    \\ 
IC 2391            &     HD 73340  & y    & n    &       & n (R)          & y        & y        & y        & y (h)    & y    \\ 
                   &     HD 74169  &      & y    & y     & y (K)          & y        & y        & y        & p (F)    & y    \\ 
                   &     HD 74168  & n    & n    &       & n (R, K)       & y        & y        &          & n (s, F) & p    \\
                   &     HD 74195  & y    & y    &       & y (R, K, *)    & n        & n        &          & n (F)    & n    \\
                   &     HD 74196  & y    & y    &       & y (R, K)       & ?        & n        &          & n (s, F) & n    \\
                   &     HD 74535  & y    & p    & y     & p (R, K)       & p        & y        & y        & n (F)    & y    \\ 
                   &     HD 74560  & y    & y    &       & y (R, K)       & ?        & n        &          & n (F)    & ?    \\
Tr 10              &     HD 75239  &      & y    &       & y (D, K)       & p        & n        &          & n (F)    & ?    \\ 
NGC 2925           &     HD 83002  & y    & y    & p     & y (D, K)       & p        &          &          & n (F)    & p    \\ 
NGC 3114           &     HD 87241  & y    & p    &       & y (B, D, K)    & p        & n        & n        & n (F)    & ?    \\ 
                   &     HD 87240  &      & n    &       & n (D, K)       & y        & y        &          & y (F)    & y    \\ 
                   &     HD 87266  & p    & p    &       & p (B, D, K)    & ?        & n        & n        & n (F)    & ?    \\ 
                   &    HD 304841  &      & p    & y     & p (D, K)       &          & y        &          & y (F)    & y    \\ 
                   &    HD 304842  &      & p    & y     & p (D, K)       & ?        & y        & y        & n (F)    & p    \\ 
                   &     HD 87405  & y    & ?    &       & ? (B, D, K)    & y        & y        & n        & n (F)    & y    \\ 
NGC 3228           &     HD 89856  & ?    & n    &       & n (B, D, K)    & ?        & y        & ?        & p (F)    & p    \\
                   &    HD 298053  &      & p    &       & p (D, K)       & n        & ?        &          & n (F)    & n    \\
vdB-Hagen 99       &     HD 92190  &      & p    & y     & p (D, K)       & p        &          &          & n (F)    & p    \\
IC 2602            &     HD 92385  & y    & y    & y     & y (R, D, K)    & ?        & y        & y        & y (s, F) & y    \\ 
                   &     HD 92664  & y    & p    & y     & y (R, K)       & y        & y        & y        & y (h)    & y    \\ 
                   &     HD 93030  & p    & ?    &       & ? (K)          & ?        &          &          & n (F)    & ?    \\
Cr 228             &    Cr 228 30  &      & p    & y     & p (K)          & n        &          &          & y (F)    & y    \\ 
                   &    HD 305451  &      & p    &       & p (K)          & p        &          &          & n (F)    & p    \\ 
NGC 3532           &     HD 96040  &      & p    & p     & p (D, K)       &          & y        &          & y (F)    & y    \\ 
                   &     HD 96729  &      & p    & ?     & ? (K)          & p        & y        &          & y (F)    & y    \\ 
                   &    HD 303821  &      & p    & ?     & ?              &          & y        &          & n (F)    & p    \\ 
Coma Ber           &    HD 108662  & y    & n    &       & n (R, K)       & y        & y        & y        & y (i)    & y    \\ 
                   &    HD 108945  & p    & y    & y     & y (R, K)       & y        & y        & y        & y (j, s, F) & y    \\ 
NGC 5460           &    HD 122983  &      & p    & y     & p (D, K)       & n        & y        &          & p (F)    & p    \\ 
                   &    HD 123183  &      & p    & y     & p (D, K)       & n        & n        &          & p (F)    & p    \\ 
                   &    HD 123225  &      & y    & y     & y (D, K)       & ?        & p        &          & n (F)    & p    \\ 
NGC 5662           &  CpD-56 6330  &      & n    &       & n (D, K)       & p        & y        &          & n (F)    & p    \\ 
                   &    HD 127866  & y    & y    &       & y (B, D, K)    & n        & n        & n        & n (F)    & n    \\ 
                   &    HD 127924  & y    & y    & y     & y (B, D, K)    & ?        & y        & n        & n (F)    & p    \\
Lower Cen-Cru (Sco OB2) & HD 103192 & y   & n    &       & n (Z)          & y        & y        & y        & y (h)    & y    \\ 
                   &    HD 112381  & p    & y    & y     & y (Z)          & y        & y        & y        & y (h)    & y    \\ 
                   &    HD 114365  & y    & y    & y     & y (Z)          & y        &          &          & n (n, s, F) & p    \\
                   &    HD 119419  & y    & y    & y     & y (Z)          & y        & y        & y        & y (h)    & y    \\ 
Upper Cen-Lup (Sco OB2) & HD 122532  & p  & n    &       & n (Z)          & y        & y        & y        & y (n)    & y    \\
                   &    HD 125823  & y    & y    & y     & y (Z)          & y        &          & y        & y (k)    & y    \\ 
                   &    HD 126759  & y    & y    & y     & y (Z)          & p        &          &          & ? (n)    & p    \\
                   &    HD 128775  & p    & y    & y     & y (Z)          & y        &          & y        & y (n, s) & y    \\ 
                   &    HD 128974  & ?    & n    &       & n (Z)          & p        &          &          & n (n, F) & ?    \\
                   &    HD 130559  & n    & n    &       & n (Z)          & y        &          & n        & y (q)    & y    \\
                   &    HD 131120  &      & y    &       & y (Z)          & ?        &          &          & n (s, F)    & ?    \\
                   &    HD 133652  & p    & y    & y     & y (Z)          & y        & y        & y        & y (h)    & y    \\ 
                   &    HD 133880  & y    & y    & y     & y (Z)          & y        & n        & y        & y (l)    & y    \\ 
                   &    HD 136347  & y    & y    & y     & y (Z, *)       & p        &          &          & n (n)    & p    \\
                   &    HD 137193  & n    & n    &       & n (Z)          & y        & y        & y        & y (n)    & y    \\
                   &    HD 139525  & n    & n    &       & n (Z)          & p        &          &          & n (n)    & p    \\
                   &    HD 143699  & y    & y    & y     & y (Z)          & ?        & n        & n        & n (k)    & ?    \\ 
Upper Sco (Sco OB2) &   HD 142301  & y    & y    & y     & y (Z)          & y        &          & y        & y (m)    & y    \\ 
                   &    HD 142990  & y    & y    & y     & y (Z)          & y        &          & y        & y (h)    & y    \\ 
                   &    HD 144334  & y    & p    & y     & y (Z)          & y        &          & y        & y (k)    & y    \\ 
                   &    HD 144661  & p    & y    & p     & y (Z)          & y        & y        & n        & n (k)    & y    \\ 
                   &    HD 145102  & y    & n    &       & n (Z)          & p        &          &          & n (n, s, F) & ?    \\
                   &    HD 145501  & y    & y    & y     & y (Z, *)       & y        & y        & ?        & y (k)    & y    \\ 
                   &    HD 146001  & y    & y    & y     & y (Z)          & y        & p        & n        & y (k)    & y    \\ 
                   &    HD 147010  & y    & y    & y     & y (Z)          & y        & y        & y        & y (n)    & y    \\ 
                   &    HD 147890  & n    & n    &       & n (Z)          & p        &          & y        & n (n)    & y    \\
                   &    HD 148199  & y    & n    &       & n (Z)          & y        & y        & ?        & y (d, h) & y    \\
                   &    HD 151525  & y    & n    &       & n (Z)          & y        & y        & y        & p (s, F) & y    \\
NGC 6087           &  CpD-57 7817  &      &      & y     & p (K)          &          & y        &          & y (F)    & y    \\ 
                   &    HD 146555  &      & y    & y     & y (D, K)       & y        & y        &          & ? (F)    & y    \\ 
NGC 6178           &    HD 149257  &      & y    &       & y (K)          & p        &          &          & n (F)    & p    \\ 
                   &    HD 149277  & y    & y    & p     & y (K)          & n        &          & ?        & y (F)    & y    \\ 
NGC 6193           &  CoD-48 11051 &      & y    & y     & y (D2, K)      & y        &          &          & y (F)    & y    \\ 
                   &  CoD-48 11059 &      &      &       & y (D2, K)      & ?        &          &          & n (F)    & ?    \\
NGC 6281           &    HD 322676  &      & p    &       & p (K)          &          &          &          & n (F)    & ?    \\ 
                   &    HD 153948  &      & y    & y     & y (D, K)       & y        & y        & y        & y (F)    & y    \\ 
NGC 6383           &  NGC 6383 26  &      &      &       & p              & ?        &          &          & n (F)    & ?    \\
                   &    HD 317857  &      & y    & ?     & ? (K)          & ?        &          &          & y (F)    & y    \\
                   &  NGC 6383 27  &      &      &       & ?              & ?        &          &          & n (F)    & ?    \\ 
NGC 6405           &    HD 318107  &      & y    & y     & y (D, K)       &          & y        & y        & y (o, F) & y    \\ 
                   &    HD 318100  &      & y    & y     & y (K)          &          & y        & y        & y (F)    & y    \\ 
                   & CoD -32 13119 &      & p    & y     & p (K)          &          & y        &          & n (F)    & p    \\ 
NGC 6475           &    HD 162305  &      & y    & y     & y (D, K)       & n        & p        &          & n (F)    & p    \\ 
                   &    HD 320764  &      & p    & y     & p (D, K)       &          & y        &          & n (F)    & p    \\ 
                   &    HD 162725  & y    & p    &       & y (B, D, K)    & y        & y        & y        & n (F)    & y    \\ 
NGC 6633           &   HD 169959A  &      & p    & ?     & ? (D, K)       & p        & y        &          & y (F)    & y    \\ 
                   &    HD 170054  & p    & y    &       & y (B, D, K)    & ?        & n        & ?        & n (F)    & ?    \\ 
IC 4725            &  BD-19 5044L  &      &      & y     & p              & n        & p        &          & n (F)    & p    \\ 
                   &    HD 170836  &      & y    & p     & p (D, K)       & ?        & y        &          & y (F)    & y    \\ 
                   &    HD 170860  &      & p    & p     & p (D, K)       & n        & y        &          & n (F)    & p    \\ 
Mel 227 = Cr 411   &    HD 190290  &      &      &       & p              & y        &          &          & y (p, F) & y    \\


\end{longtable}

%
%
%
%

\begin{tabular}{ll}
\multicolumn{2}{l}{Notes on individual stars (* in ``member?'' or ``Ap?'' column):} \\
HD 19805     &   Composition normal, no field, not Ap star (private communication, J. Silvester) \\
HD 23387A    &   Composition normal, no field, not Ap star (private communication, J. Silvester) \\
HD 37022     &   Tycho-2 $\mu_\delta$ inconsistent with Hipparcos
value and with membership of this Trapezium star in Ori OB1d. \\
             &   Accept Hipparcos value. \\
CpD -20 1640 &   This star is Cox 40. Simbad identification and coordinates of NGC 2287 40 are incorrect. \\
HD 56343     &   Parallax and proper motions agree with cluster means,
but Baumgardt et al 00 regard this star as \\
             &   non-member. We accept it as member. \\
HD 62376     &   Carrier et al. (\cite{carrier99}) and Baumgardt et al. (\cite{baumgardt00}) accept star
as probable member. Because of 5$\sigma$ discrepancy in \\
             &   $\mu_\delta$ we consider it a non-member. \\
HD 65712     &   Star seems to be outside field considered by Dias et al 2001. \\
CpD -60 944B &   Dias et al 2001 consider this star non-member, but both
proper motions differ from cluster mean by less than \\
             &   2$\sigma$. Accept as probable member. \\
HD 65949     &    Both Hipparcos and Tycho-2 proper motions have large
uncertainties, but are consistent with probable \\
             &  membership. Disregard small membership probability assigned by Dias et al 2001. \\
HD 74195     &   Hipparcos and Tycho-2 $\mu_\delta$'s inconsistent. Accept Tycho-2 value which is consistent with membership. \\
HD 136347    &   Tycho-2 $\mu$'s both completely different from Hipparcos values. Accept Hipparcos values and membership. \\
HD 145501    &   Membership assessment for $\nu$ Sco CD based on $\pi$ and $\mu$'s for $\nu$ Sco AB. \\
HD 190290    &  Cluster mean motions unknown for Melotte 227. 
\end{tabular}

\vspace*{2mm}
\begin{tabular}{ll}
\multicolumn{2}{l}{References for ``member?'' column:} \\ 
(B)  &  Baumgardt et al. (\cite{baumgardt00}) \\ 
(Br) &  Brown et al. (\cite{brown99}) \\ 
(C)  &  Carrier et al. (\cite{carrier99}) \\ 
(D)  &  Dias et al. (\cite{dias01}) \\ 
(D2) &  Dias et al. (\cite{dias02}) \\
(K)  &  Kharchenko et al. (\cite{kharchenko05}) \\
(M)  &  Marschall et al. (\cite{marschall82}) \\
(R)  &  Robichon et al. (\cite{robichon99}) \\
(S)  &  Sabogal-Mart\'\i nez et al. (\cite{sabogal01}) \\
(T)  &  Tian et al. (\cite{tian96}) \\
(Z)  &  de Zeeuw et al. (\cite{dezeeuw99}) 
\end{tabular}

\vspace*{2mm}
\begin{tabular}{ll}
\multicolumn{2}{l}{References for ``mag fld'' column:} \\
(F)  &  Bagnulo et al. (\cite{Bagea06}), FORS1 data from Paper I \\
(a)  &  Bychkov et al. (\cite{Bycetal03})  \\
(b)  &  Brown et al. (\cite{Broetal85}\\
(c)  &  Borra (\cite{Borra81}) \\
(d)  &  Bohlender et al. (\cite{Bohetal87}) \\
(e)  &  Wade et al (\cite{WadFulDon06}) \\
(g)  &  Thompson \& Landstreet (\cite{ThoLan85}) \\
(h)  &  Bohlender et al. (\cite{Bohetal93}) \\
(i)  &  Preston et al. (\cite{PreSte69}) \\
(j)  &  Shorlin et al. (\cite{Shoetal02}) \\
(k)  &  Borra et al. (\cite{Boretal83}) \\
(l)  &  Landstreet (\cite{Lan90}) \\
(m)  &  Landstreet et al. (\cite{Lanetal79}) \\
(n)  &  Thompson et al. (\cite{Thoetal87}) \\
(o)  &  Mathys et al. (\cite{Matetal97}) \\
(p)  &  Hubrig et al. (\cite{Hubetal04}) \\
(q)  &  Babcock (\cite{Bab58}) \\
(r)  &  unpublished Musicos data, courtesy J. Silvester \\
(s)  &  Kochukhov \& Bagnulo (\cite{KocBag06})
\end{tabular}

\begin{longtable}[c]{lrrlrrrrr} 
 \caption{\label{cl_stell_params_magmeas.tab} Physical properties of stars that 
 are probable open cluster members, probable Ap stars, and have 
 magnetic field measurements. 
 }\\ 
\hline 
\hline \\ 
 Cluster  &   $\log t$ & true DM &    Star      & $\log T_e$ & $\log L/L_\odot$ &       $M/M_\odot$   & fractional age       & $B_{\rm rms}$ \\ 
\hline  
\endfirsthead 
\hline \\ 
Cluster  &   $\log t$ & true DM &    Star      & $\log T_e$ & $\log L/L_\odot$ &       $M/M_\odot$   & fractional age       & $B_{\rm rms}$ \\ 
\hline 
\endhead 
\hline 
\endfoot 
\hline 
\endlastfoot 
NGC 1039      &   8.30 $\pm$   0.15 &  8.489  &   HD 16605      &  4.025  &         1.65      &   2.55 $\pm$   0.15 &    0.37 $\pm$   0.15 &    1700 \\ 
$\alpha$ Per  &   7.93 $\pm$   0.10 &   6.40  &   HD 21699      &  4.158  &         2.78      &   4.60 $\pm$   0.20 &    0.70 $\pm$   0.21 &     565 \\ 
NGC 1662      &   8.64 $\pm$   0.06 &   8.20  &   HD 30598      &  3.939  &         1.90      &   2.65 $\pm$   0.15 &    0.91 $\pm$   0.21 &     190 \\ 
Ori OB1a      &   7.00 $\pm$   0.10 &   7.63  &   HD 35298      &  4.201  &         2.36      &   4.25 $\pm$   0.25 &    0.07 $\pm$   0.02 &    2280 \\ 
              &                 &         &   HD 35502      &  4.209  &         2.66      &   4.80 $\pm$   0.20 &    0.09 $\pm$   0.03 &    1520 \\ 
Ori OB1b      &   6.55 $\pm$   0.15 &   8.37  &   HD 36485      &  4.290  &         3.27      &   6.80 $\pm$   0.30 &    0.07 $\pm$   0.03 &    3220 \\ 
              &                 &         &   HD 36526      &  4.212  &         2.54      &   4.65 $\pm$   0.20 &    0.03 $\pm$   0.01 &    2230 \\ 
              &                 &         &   HD 37140      &  4.190  &         2.61      &   4.60 $\pm$   0.20 &    0.03 $\pm$   0.01 &     580 \\ 
              &                 &         &   HD 37479      &  4.382  &         3.55      &   8.95 $\pm$   0.40 &    0.12 $\pm$   0.05 &    1910 \\ 
              &                 &         &   HD 37633      &  4.121  &         2.14      &   3.50 $\pm$   0.20 &    0.02 $\pm$   0.01 &     285 \\ 
              &                 &         &   HD 37776      &  4.369  &         3.54      &   8.80 $\pm$   0.40 &    0.11 $\pm$   0.04 &    1260 \\ 
Ori OB1c      &   6.66 $\pm$   0.20 &   8.52  &   HD 36540      &  4.196  &         2.77      &   4.80 $\pm$   0.30 &    0.04 $\pm$   0.02 &     275 \\ 
              &                 &         &   HD 37058      &  4.312  &         3.26      &   7.05 $\pm$   0.30 &    0.09 $\pm$   0.05 &     710 \\ 
              &                 &         &   HD 37642      &  4.164  &         2.56      &   4.25 $\pm$   0.20 &    0.03 $\pm$   0.02 &    2140 \\ 
Ori OB1d      &   6.00 $\pm$   0.20 &   8.52  &   HD 37022      &  4.595  &         4.95      &  45.00 $\pm$   5.00 &    0.33 $\pm$   0.16 &     205 \\ 
NGC 2169      &   6.97 $\pm$   0.10 &  10.11  &   NGC 2169 12   &  4.140  &         2.20      &   3.65 $\pm$   0.15 &    0.04 $\pm$   0.01 &    3410 \\ 
NGC 2232      &   7.55 $\pm$   0.10 &   7.56  &   HD 45583      &  4.104  &         2.04      &   3.30 $\pm$   0.15 &    0.13 $\pm$   0.04 &    2730 \\ 
NGC 2244      &   6.40 $\pm$   0.10 &  10.80  &   NGC 2244 334  &  4.204  &         2.35      &   4.30 $\pm$   0.30 &    0.02 $\pm$   0.01 &    9515 \\ 
NGC 2281      &   8.63 $\pm$   0.05 &  8.733  &   HD 49040      &  3.944  &         1.93      &   2.65 $\pm$   0.15 &    0.89 $\pm$   0.19 &     160 \\ 
NGC 2287      &   8.32 $\pm$   0.12 &   9.20  &   CPD-20 1640   &  3.908  &         1.16      &   1.85 $\pm$   0.10 &    0.16 $\pm$   0.05 &     155 \\ 
              &                 &         &   HD 49299      &  3.987  &         1.60      &   2.40 $\pm$   0.10 &    0.33 $\pm$   0.11 &    2720 \\ 
CMa OB2       &   7.05 $\pm$   0.20 &   8.37  &   HD 51088      &  4.097  &         2.23      &   3.41 $\pm$   0.25 &    0.04 $\pm$   0.02 &      90 \\ 
NGC 2323      &   8.15 $\pm$   0.07 &   9.84  &   HD 52965      &  4.090  &         2.64      &   4.05 $\pm$   0.20 &    0.86 $\pm$   0.21 &      65 \\ 
Cr 132        &   7.30 $\pm$   0.20 &   8.37  &   HD 56343      &  4.061  &         2.03      &   3.05 $\pm$   0.15 &    0.06 $\pm$   0.03 &    3610 \\ 
NGC 2422      &   8.08 $\pm$   0.11 &   8.48  &   HD 61045      &  4.114  &         2.47      &   3.85 $\pm$   0.20 &    0.65 $\pm$   0.20 &     430 \\ 
NGC 2451A     &   7.70 $\pm$   0.10 &   6.38  &   HD 63401      &  4.130  &         2.25      &   3.70 $\pm$   0.20 &    0.24 $\pm$   0.07 &     365 \\ 
NGC 2516      &   8.15 $\pm$   0.10 &   7.77  &   HD 65712      &  3.996  &         1.41      &   2.30 $\pm$   0.10 &    0.20 $\pm$   0.06 &     865 \\ 
              &                 &         &   CPD -60 944A  &  4.100  &         2.10      &   3.30 $\pm$   0.15 &    0.52 $\pm$   0.15 &     250 \\ 
              &                 &         &   CPD -60 944B  &  4.107  &         2.03      &   3.25 $\pm$   0.20 &    0.50 $\pm$   0.15 &     120 \\ 
              &                 &         &   CPD-60 978    &  4.072  &         1.88      &   3.06 $\pm$   0.15 &    0.43 $\pm$   0.13 &      85 \\ 
              &                 &         &   HD 65987      &  4.100  &         2.32      &   3.60 $\pm$   0.20 &    0.64 $\pm$   0.19 &     540 \\ 
              &                 &         &   HD 66295      &  4.045  &         1.65      &   2.60 $\pm$   0.15 &    0.28 $\pm$   0.08 &     490 \\ 
              &                 &         &   HD 66318      &  3.959  &         1.31      &   2.05 $\pm$   0.10 &    0.15 $\pm$   0.04 &    4385 \\ 
NGC 2546      &    8.20 $\pm$   0.05 &   9.82  &   NGC 2546 201  &  4.079  &         2.40      &   3.70 $\pm$   0.40 &    0.77 $\pm$   0.17 &     595 \\ 
              &                 &         &   NGC 2546 197  &  4.000  &         1.62      &   2.40 $\pm$   0.30 &    0.25 $\pm$   0.05 &      20 \\ 
IC 2391       &   7.70 $\pm$   0.15 &   5.82  &   HD 74169      &  4.009  &         1.43      &   2.35 $\pm$   0.10 &    0.08 $\pm$   0.03 &     130 \\ 
              &                 &         &   HD 74535      &  4.133  &         2.39      &   3.85 $\pm$   0.15 &    0.27 $\pm$   0.11 &      95 \\ 
NGC 2925      &   8.17 $\pm$   0.13 &   9.44  &   HD 83002      &  4.093  &         2.41      &   3.70 $\pm$   0.25 &    0.72 $\pm$   0.25 &      20 \\ 
NGC 3114      &   8.13 $\pm$   0.15 &  10.01  &   HD 304841     &  4.090  &         2.17      &   3.40 $\pm$   0.15 &    0.53 $\pm$   0.21 &     335 \\ 
              &                 &         &   HD 304842     &  4.097  &         2.29      &   3.55 $\pm$   0.15 &    0.59 $\pm$   0.23 &      20 \\ 
vdB-H 99      &   7.80 $\pm$   0.20 &   8.63  &   HD 92190      &  4.104  &         2.28      &   3.55 $\pm$   0.15 &    0.28 $\pm$   0.14 &      10 \\ 
IC 2602       &   7.65 $\pm$   0.20 &   5.91  &   HD 92385      &  4.045  &         1.75      &   2.70 $\pm$   0.15 &    0.10 $\pm$   0.05 &     440 \\ 
              &                 &         &   HD 92664      &  4.152  &         2.48      &   4.05 $\pm$   0.20 &    0.27 $\pm$   0.13 &     810 \\ 
Cr 228        &   6.60 $\pm$   0.20 &  12.77  &   Cr 228 30     &  4.362  &         3.38      &   8.00 $\pm$   0.40 &    0.11 $\pm$   0.05 &     560 \\ 
NGC 3532      &   8.45 $\pm$   0.10 &   8.47  &   HD 96040      &  4.025  &         1.47      &   2.40 $\pm$   0.15 &    0.45 $\pm$   0.13 &     260 \\ 
Coma Ber      &   8.70 $\pm$   0.10 &   4.70  &   HD 108945     &  3.944  &         1.60      &   2.30 $\pm$   0.15 &    0.71 $\pm$   0.21 &     110 \\ 
NGC 5460      &   8.17 $\pm$   0.10 &   9.34  &   HD 122983     &  4.029  &         1.82      &   2.70 $\pm$   0.15 &    0.32 $\pm$   0.10 &     170 \\ 
              &                 &         &   HD 123183     &  4.000  &         1.71      &   2.50 $\pm$   0.10 &    0.26 $\pm$   0.08 &     410 \\ 
              &                 &         &   HD 123225     &  4.090  &         2.38      &   3.65 $\pm$   0.15 &    0.70 $\pm$   0.21 &      20 \\ 
NGC 5662      &   7.82 $\pm$   0.18 &  10.08  &   HD 127924     &  4.121  &         2.58      &   4.05 $\pm$   0.20 &    0.40 $\pm$   0.18 &      10 \\ 
Lower Cen Cru &   7.15 $\pm$   0.15 &   5.36  &   HD 112381     &  4.000  &         1.53      &   2.35 $\pm$   0.10 &    0.02 $\pm$   0.01 &    3400 \\ 
              &                 &         &   HD 114365     &  4.072  &         1.83      &   2.85 $\pm$   0.15 &    0.04 $\pm$   0.01 &     290 \\ 
              &                 &         &   HD 119419     &  4.045  &         1.62      &   2.60 $\pm$   0.15 &    0.03 $\pm$   0.01 &    1430 \\ 
Upper Cen Lup &   7.20 $\pm$   0.10 &   5.73  &   HD 125823     &  4.283  &         3.16      &   6.40 $\pm$   0.30 &    0.27 $\pm$   0.08 &     320 \\ 
              &                 &         &   HD 126759     &  4.086  &         1.87      &   2.95 $\pm$   0.15 &    0.04 $\pm$   0.01 &     335 \\ 
              &                 &         &   HD 128775     &  4.079  &         1.76      &   2.85 $\pm$   0.15 &    0.04 $\pm$   0.01 &     345 \\ 
              &                 &         &   HD 133652     &  4.114  &         2.09      &   3.35 $\pm$   0.15 &    0.06 $\pm$   0.02 &    1120 \\ 
              &                 &         &   HD 133880     &  4.079  &         2.10      &   3.20 $\pm$   0.15 &    0.05 $\pm$   0.02 &    2300 \\ 
              &                 &         &   HD 136347     &  4.056  &         1.78      &   2.75 $\pm$   0.15 &    0.04 $\pm$   0.01 &     225 \\ 
              &                 &         &   HD 143699     &  4.199  &         2.72      &   4.80 $\pm$   0.20 &    0.14 $\pm$   0.04 &     165 \\ 
Upper Sco     &   6.70 $\pm$   0.10 &   5.81  &   HD 142301     &  4.204  &         2.52      &   4.60 $\pm$   0.20 &    0.04 $\pm$   0.01 &    2385 \\ 
              &                 &         &   HD 142990     &  4.258  &         2.79      &   5.40 $\pm$   0.25 &    0.06 $\pm$   0.02 &    1080 \\ 
              &                 &         &   HD 144334     &  4.212  &         2.50      &   4.60 $\pm$   0.20 &    0.04 $\pm$   0.01 &     715 \\ 
              &                 &         &   HD 145501     &  4.176  &         2.53      &   4.30 $\pm$   0.20 &    0.04 $\pm$   0.01 &    1385 \\ 
              &                 &         &   HD 146001     &  4.146  &         2.41      &   3.90 $\pm$   0.20 &    0.03 $\pm$   0.01 &     650 \\ 
              &                 &         &   HD 147010     &  4.111  &         1.92      &   3.15 $\pm$   0.20 &    0.02 $\pm$   0.00 &    4825 \\ 
NGC 6087      &   7.90 $\pm$   0.10 &  10.29  &   CPD-57 7817   &  4.079  &         2.25      &   3.35 $\pm$   0.20 &    0.30 $\pm$   0.09 &     610 \\ 
              &                 &         &   HD 146555     &  4.107  &         2.19      &   3.45 $\pm$   0.20 &    0.33 $\pm$   0.10 &     270 \\ 
NGC 6178      &   7.15 $\pm$   0.22 &  10.03  &   HD 149257     &  4.406  &         3.82      &  10.30 $\pm$   0.60 &    0.61 $\pm$   0.33 &     160 \\ 
              &                 &         &   HD 149277     &  4.348  &         3.71      &   8.75 $\pm$   0.40 &    0.45 $\pm$   0.24 &    2435 \\ 
NGC 6193      &   6.69 $\pm$   0.20 &  10.31  &   CD-48 11051   &  4.301  &         3.32      &   7.00 $\pm$   0.30 &    0.10 $\pm$   0.05 &    2400 \\ 
NGC 6281      &   8.45 $\pm$   0.10 &   8.86  &   HD 153948     &  4.025  &         1.86      &   2.70 $\pm$   0.15 &    0.62 $\pm$   0.18 &     195 \\ 
NGC 6405      &   7.80 $\pm$   0.15 &   8.44  &   HD 318107     &  4.072  &         1.92      &   2.95 $\pm$   0.15 &    0.17 $\pm$   0.07 &    4820 \\ 
              &                 &         &   HD 318100     &  4.025  &         1.64      &   2.50 $\pm$   0.10 &    0.11 $\pm$   0.04 &     345 \\ 
              &                 &         &   CoD -32 13119 &  3.892  &         1.06      &   1.75 $\pm$   0.10 &    0.04 $\pm$   0.02 &      35 \\ 
NGC 6475      &   8.47 $\pm$   0.13 &   7.24  &   HD 162305     &  4.004  &         1.82      &   2.65 $\pm$   0.15 &    0.61 $\pm$   0.22 &      85 \\ 
              &                 &         &   HD 162725     &  3.982  &         2.36      &   3.30 $\pm$   0.20 &    1.08 $\pm$   0.38 &      60 \\ 
NGC 6633      &   8.75 $\pm$   0.05 &   7.93  &   HD 169842     &  3.924  &         1.62      &   2.35 $\pm$   0.10 &    0.85 $\pm$   0.18 &     370 \\ 
IC 4725       &   8.02 $\pm$   0.08 &  10.44  &   BD-19 5044L   &  4.107  &         2.25      &   3.55 $\pm$   0.15 &    0.46 $\pm$   0.12 &     235 \\ 
              &                 &         &   HD 170836     &  4.133  &         2.80      &   4.55 $\pm$   0.20 &    0.84 $\pm$   0.22 &     505 \\ 
              &                 &         &   HD 170860     &  4.137  &         2.63      &   4.25 $\pm$   0.20 &    0.71 $\pm$   0.19 &      40 \\ 
\hline 
\hline 
\end{longtable} 

\end{document}